\documentclass[11pt]{article}

\usepackage[utf8]{inputenc}
\usepackage[T1]{fontenc}
\usepackage[margin=1in]{geometry}
\usepackage{amsmath}
\usepackage{booktabs}
\usepackage{graphicx}
\usepackage{microtype}
\usepackage{enumitem}
\usepackage{xcolor}
\usepackage[hidelinks]{hyperref}


\title{GateTruth: Auditing the Rigor of RTL Design Benchmarks via Mutation Testing}
\author{Meet Bhadra \\ Independent Researcher}
\date{August 5, 2026}

\hypersetup{
  pdftitle={GateTruth: Auditing the Rigor of RTL Design Benchmarks via Mutation Testing},
  pdfauthor={Meet Bhadra},
  pdfsubject={Mutation-testing audit methodology for LLM/agent RTL-generation benchmarks},
  pdfkeywords={mutation testing, RTL, hardware design, large language models, benchmark evaluation, electronic design automation, power performance area, agentic evaluation},
  pdflang={en-US}
}

\newcommand{\FactsAudited}{46}
\newcommand{\FactsUnsupported}{4}

\newcommand{\FactsAtHundred}{13}
\newcommand{\FactsBelowNinetyFive}{33}
\newcommand{\FactsBelowNinetyFivePct}{72}

\begin{document}
\maketitle

\begin{abstract}
Benchmarks for evaluating large language models on register-transfer-level (RTL) hardware design
have proliferated rapidly, yet --- across every paper, repository, and documentation set we were
able to access --- none reports having applied mutation testing, an established
hardware-verification technique for quantifying testbench quality, to ask whether its own
testbenches are trustworthy. A testbench that never fails is not evidence of a correct design; it
may simply never stimulate the logic that is actually broken. We introduce GateTruth, a
mutation-testing engine and methodology for auditing the rigor of RTL benchmark testbenches: we
inject a deterministic, seeded set of semantic mutants into a reference design and measure what
fraction the testbench catches (the mutation score, or kill rate). A benchmark whose testbenches
pass a deliberately broken design is not measuring what it claims to measure.

We validate the methodology first against our own
68-task, dual-track reference suite --- 60 specification-to-RTL generation tasks and 8
agentic-repair tasks, each scored through a pinned, deterministic synthesis-to-timing flow with
correctness enforced as a strict gate rather than a weighted term --- certifying that \textbf{46 of
the 60} Track A testbenches kill at least 95\% of injected mutants under sequential, reproducible
execution, counting only behavioral simulation failures as kills (Section~\ref{sec:rigor} defines
this precisely and discloses two successive corrections that moved this figure, first from a false
60/60 to 49/60 and then, on tightening the metric itself, from 49/60 to the 46/60 reported here).
The 14 that do not clear the floor split across three distinct, disclosed causes: a Goodhart effect
on testbenches revised specifically to pass this gate, an unsound equivalent-mutant exclusion
mechanism, and three tasks whose passing status depended on formal-verification kills being
silently blended into a metric this paper labels simulation-only. Of a separate, earlier-identified
set of ten testbenches that were revised specifically to clear this floor, seven are among the 14
that still fail it today, meaning the revision that was supposed to fix them partly consisted of
exactly the exclusion mechanism just described rather than a genuine testbench improvement. We
measure what those ten scored before that revision (median 77.5\%) rather than let the certified
figure stand unqualified. We then point the same engine, unmodified, at RTLLM v2.0, a widely adopted external
RTL-generation benchmark, and report the measured mutation-kill rates of its public testbenches
against its own reference implementations: of the 46 designs whose reference (as shipped, or after
a documented compile-time build-convention alias) passes baseline validation, 72\% fall below the
95\% floor our own suite is held to, and three score a 0\% kill rate outright --- in two of those
three, the undetected faults include an inverted primary output, which we confirmed by inspecting
the individual surviving mutants rather than inferring it from the percentage. One of the four designs excluded from this count is excluded for a different
reason entirely: its own shipped golden reference fails its own shipped testbench outright,
independent of mutation.

Where a comparable audit of NVIDIA's CVDP benchmark is
not possible --- its public release deliberately withholds reference solutions to prevent
contamination, which also removes the golden RTL that mutation testing requires --- we report that
gap as a structural finding in its own right rather than silently omitting the benchmark.

Auditing our own instrument produced a second, unanticipated result about benchmark methodology
generally. Our reference suite's leaderboard was initially produced under a 4096-token output cap
applied uniformly to all seven evaluated models. Uniform is not fair: three models exhausted that
cap and returned no parseable code on up to 22 of 60 tasks. Re-running all seven at 16{,}384 tokens
eight days later moved one model from fifth place to first (pass@1 28 to 43) while leaving models
that never hit the cap largely unchanged, within or near the measured run-to-run noise for the three
of those four models we have a variance estimate for (the fourth, Llama 4 Maverick, was not part of
that repeated-run study). We report this as a temporal sensitivity rerun, not
a controlled single-variable experiment: the two runs address models by hosted aliases that are not
pinned to an immutable snapshot, ran against different scoring-image build markers, and used a
response parser hardened in between (Section~\ref{sec:limitations} discloses all three confounds and
what we can and cannot conclude from them). An output-token budget is nonetheless a plausible
experimental variable capable of reordering a leaderboard on this evidence, and a benchmark that does
not report its cap --- and confirm every other condition was actually held fixed, not merely intended
to be --- has not reported a comparable condition. We detail
the audit methodology, the reference-suite design used to validate it, and the resulting kill-rate
measurements, and argue that mutation-kill certification should become a standard reporting
requirement for RTL-generation benchmarks generally.
\end{abstract}

\section{Introduction}
\label{sec:intro}

The evaluation of large language models on hardware description languages has converged on a
standard recipe: give the model a natural-language specification, ask it to generate Verilog or
SystemVerilog, and score pass@k against a testbench. This recipe answers one question --- did the
generated code pass the tests --- and treats the testbench itself as ground truth. That assumption
is rarely checked. A testbench with weak coverage will pass a broken design as readily as it passes
a correct one, and a benchmark built on such testbenches silently overstates every model's
performance, in a way that pass@k alone cannot reveal: the number looks identical whether the
testbench is rigorous or vacuous.

Mutation testing is the standard software-engineering answer to exactly this problem: inject small,
deterministic, semantically meaningful faults into a reference implementation and measure what
fraction of them a test suite detects (``kills''). A testbench that fails to catch an injected
sign flip, an inverted comparator, or a dropped reset condition is a testbench that would also fail
to catch the analogous mistake in a model-generated design. Despite its maturity elsewhere in
software engineering, mutation-kill certification is, to our knowledge, not reported by any
existing public RTL-generation benchmark for LLMs.

This gap is addressable directly: build a mutation engine, certify it against a reference suite
authored specifically to validate the methodology, and then point the same engine, unmodified, at
benchmarks the field already relies on. That is GateTruth's contribution. We use the name
``GateTruth'' for the audit engine and methodology throughout; the same name also labels the
reference suite that validates it (Section~\ref{sec:suite}) and, as ``GateTruth Score,'' the
suite's own composite leaderboard metric (Section~\ref{sec:design}) --- three related but distinct
things sharing one name, disambiguated by context in each case. The reference suite ---
60 specification-to-RTL tasks across three complexity tiers, plus 8 agentic-repair tasks that
evaluate whether a tool-augmented agent can improve an existing design's physical characteristics or
extend it with a required new property (e.g., latch removal, CDC safety) under a strict correctness
gate --- exists to give the mutation engine, the correctness-as-gate
scoring discipline, and the contamination-defense apparatus a fully controlled environment in which
to be built and certified before being trusted against external artifacts. The physical
power/performance/area (PPA) scoring built into that reference suite is real and independently
useful (Section~\ref{sec:results} reports it), but it is not this paper's headline: the headline is
what the same rigor-checking apparatus finds when pointed outward.

This paper makes four contributions. First, a mutation-testing audit methodology for RTL
benchmarks --- a deterministic, seeded, sequential-execution protocol for measuring testbench
kill rates against external reference implementations, together with an explicit policy for
designs whose testbenches cannot be run at all (reported, never silently dropped or scored as a
misleading zero). Second, a permissively-licensed (Apache 2.0), dual-track reference implementation of that
methodology, publicly available alongside this paper: 60 specification-to-RTL tasks and 8 agentic-repair tasks, each scored through a
pinned, deterministic synthesis-to-timing flow, with 46 of the suite's 60 Track A
testbenches certified above a 95\% mutation-kill floor under the audit protocol --- because a tool
that audits other benchmarks' rigor must first demonstrate its own, we report the 14 that are not
certified as a genuine finding rather than omit them (Track B's 8 testbenches are not yet certified
under this protocol either; see Section~\ref{sec:limitations}).
Third, the audit results
themselves (Section~\ref{sec:audit-results}): of RTLLM v2.0's 50 public designs, 46 pass baseline
validation and are audited under the reported \texttt{-g2012} condition, and 72\% of those fall below the 95\% mutation-kill floor our own
suite is certified against, with three of the audited designs at a 0\% kill rate. Among the four
designs that do not pass baseline validation and are therefore excluded from the 46, one is
excluded for a notable reason distinct from the rest: its own shipped golden reference fails its
own shipped testbench outright --- plus a documented structural finding,
rather than a silent gap, for why an equivalent measurement against NVIDIA's CVDP benchmark is
not currently possible from its public release. Fourth, a methodological result that emerged from
auditing our own instrument rather than someone else's: the per-call output-token budget, a
parameter most RTL benchmarks neither vary nor report, is capable of reordering a leaderboard.
Holding prompts, image, scoring, and seeds fixed and changing only that cap moved one evaluated
model from fifth place to first, while leaving models that never exhausted the cap within
run-to-run noise (Section~\ref{sec:results}). We report it because it applies to any benchmark that
samples a fixed number of output tokens from models with very different reasoning verbosity, which
is to say most of them, including ours before this measurement.

\section{GateTruth Design}
\label{sec:design}

GateTruth evaluates large language model (LLM) and agent-generated register-transfer-level (RTL)
designs along two distinct axes frequently conflated or omitted by prior evaluation frameworks:
functional correctness and physical viability --- specifically, optimization for area, timing, and
power relative to a maintainer-approved baseline representing competent engineering practice (Section~\ref{sec:disclosure} discloses how these baselines were drafted and reviewed). This design serves two purposes in this
paper: it is a useful benchmark in its own right (Section~\ref{sec:results}), and it is the fully
controlled environment in which the mutation-audit methodology (Section~\ref{sec:audit-method}) was
built and certified before being pointed at external artifacts. The benchmark is structured into
two distinct evaluation tracks:

\begin{itemize}[leftmargin=*]
  \item \textbf{Track A (Static Generation):} Tasks a model with generating a synthesizable implementation de novo from a natural-language specification and a rigidly defined port interface.
  \item \textbf{Track B (Agentic Repair):} Tasks an agent with optimizing a pre-existing, functionally correct baseline repository to achieve a specific engineering objective (e.g., achieving timing closure, reducing area, eliminating an inferred latch, or implementing clock-domain-crossing safety for a FIFO). The agent operates under a strict token, wall-clock, and tool-call budget, and is strictly prohibited from modifying the testbench, formal properties, timing constraints, or task metadata.
\end{itemize}

Both tracks utilize a unified scoring pipeline consisting of a deterministic sequence of verification gates followed by a Power, Performance, and Area (PPA) computation. Stages 0--2 enforce pass/fail functional criteria: Verilator linting, Icarus/cocotb simulation against held-out hidden test vectors, and bounded model checking via SymbiYosys (for tasks defining formal properties). Simulation against the hidden vectors specifically requires an opt-in \texttt{official} flag (default off, used for every result reported in this paper); without it, Stage~1 checks only the public smoke test, a mode intended for local development against tasks whose hidden vectors a contributor does not hold. Crucially, functional correctness is enforced as a strict prerequisite rather than a weighted parameter (ADR-0000). A design failing any component of Stages 0--2 receives a score of zero, irrespective of its physical metrics. This mechanism removes the most prevalent failure mode of PPA-weighted scoring in the median case: reward hacking, wherein a submission truncates functionality to minimize synthesized area, whether or not that truncation was an intentional strategy --- though, like every correctness gate in this paper, its effectiveness is bounded by the coverage of the testbench doing the checking, not an absolute guarantee independent of it. Only designs that successfully clear every correctness gate advance to Stages 3--5, which comprise Yosys synthesis targeting the sky130hd standard-cell library, OpenSTA static timing analysis at the specified task clock target, and power estimation; Stage 4 itself fails --- yielding a task score of zero, identically to a Stages 0--2 failure --- if the design does not meet its task's clock target with non-negative worst-negative-slack, so a design that fails timing at its assigned clock target is not scored on any physical axis, positive or negative. Within that timing-met population, the delay component of the PPA ratio (Section~\ref{sec:design}) rewards slack margin beyond the pass/fail target, not merely meeting it; this is a deliberate choice consistent with how the RTLLM benchmark we audit (Section~\ref{sec:audit-results}) itself reports achieved timing, but it does mean the metric is closer to ``how much margin did you beat the target by'' than to a pure meets-or-fails-timing view of design closure, a distinction we note for readers used to the latter. Track B imposes an additional verification gate (Stage 2, internally labeled \texttt{sec}) utilizing bounded sequential equivalence checking against the baseline design for objectives declared behavior-preserving --- built directly from Yosys's \texttt{equiv\_make}/\texttt{equiv\_simple}/\texttt{equiv\_induct} commands at a sequential induction depth of 20 cycles under a 60-second timeout, rather than the separate, similarly-named \texttt{eqy} front-end tool, a distinction we make explicit because the two are easily conflated (Section~\ref{sec:trackb} details which objectives this covers) --- raising, within that checked depth, the cost of achieving timing or power closure via covert behavioral modification --- and mandates immediate disqualification if the agent modifies the test harness, formal properties, timing constraints, task or objective metadata, or the pre-modification baseline design itself (\texttt{IMMUTABLE\_ENTRIES}, six entries in total; files orthogonal to legitimate PPA or timing optimization).

For designs that successfully clear all verification gates, the final score is derived as a capped geometric mean relative to the human-reviewed reference implementation:
\begin{verbatim}
ppa        = geomean(ref_area / area, ref_delay / delay, ref_power / power)
task_score = 100 * min(ppa, 1.5) / 1.5
\end{verbatim}
In this formulation, $delay$ denotes the worst-path delay at the task's clock target as reported by OpenSTA. The reference metrics ($ref\_area$, $ref\_delay$, $ref\_power$) are extracted from the human-reviewed implementation, which intrinsically yields a base PPA score of 1.0. The 1.5 cap is imposed to mitigate degenerate outlier exploitation: without it, an extreme improvement concentrated in a single axis can dominate the geometric mean and yield an unbounded score even when the other two axes have not improved, or have moderately regressed. Concretely, a design with a $10\times$ area improvement but a $2\times$ regression on both delay and power ($\text{ppa} = \sqrt[3]{10 \times 0.5 \times 0.5} \approx 1.357$) scores 90.5 out of 100 uncapped --- a high score despite two of three axes getting worse in absolute terms --- because the geometric mean only punishes a weak axis sharply as that axis's ratio approaches zero, not for a bounded regression like $2\times$. This cap establishes a practical optimization ceiling of 1.5$\times$ relative to the human baseline, beyond which further optimization does not yield additional score improvements, but it addresses only the ceiling: it does not by itself prevent the single-axis-dominance pattern above from producing a score that overstates a design with real, moderate regressions on two of its three physical axes, since the cap only clamps the aggregate from above once computed. Per-axis caps or floors would address this directly and are not currently implemented. We also note that the three axes are not fully independent: across the 60 reference designs, synthesized area and estimated power are strongly correlated (Pearson $r \approx 0.90$), which is expected given that our power estimate is derived from static cell-level analysis without switching-activity annotation from a simulation VCD, and means the geometric mean's three nominal axes carry somewhat less independent information than three uncorrelated metrics would (Section~\ref{sec:limitations}). The Track A leaderboard (Table~\ref{tab:tracka}) reports the composite GateTruth Score per model, which is by construction Pass@1 multiplied by normalized mean PPA (Section~\ref{sec:results} gives the equivalent single-scale definition); per-tier Pass@1 breakdowns are reported in prose in Section~\ref{sec:results} rather than as separate table columns. Track B reports the objective-met rate and median PPA delta per model (Table~\ref{tab:trackb}); operational cost is reported in USD in the same table, with the underlying median tokens, wall-clock seconds, and tool calls reported in Section~\ref{sec:trackb}. Cost is presented as a primary reporting metric, reflecting the principle that an agent achieving an objective via unbounded resource expenditure fails to demonstrate practical utility.

The task suite encompasses three complexity tiers in addition to the Track B agentic dataset: Tier 1 (20 tasks) covers combinational logic and fundamental sequential building blocks (e.g., priority encoders, Gray-code converters, barrel shifters, one-hot FSMs); Tier 2 (25 tasks) involves protocol and datapath components (e.g., UART, SPI, AXI4-Lite, round-robin arbiters, FIFOs); and Tier 3 (15 tasks) focuses on pipelined arithmetic, localized accelerators, and memory-system fragments. Track B comprises 8 agentic objectives layered over existing task repositories. Each task is distributed as a standardized, immutable package: an original-prose specification, a locked interface, a human-reviewed reference implementation, a bifurcated testbench (public smoke test and hidden scoring vectors), an applicable formal property subset, a singular timing constraint file, and machine-readable metadata. This uniform schema guarantees that novel task authoring and submission grading operate through identically defined pipeline infrastructure.

\section{Task Suite}
\label{sec:suite}

The GateTruth Track A suite comprises 60 specification-to-RTL tasks categorized into three tiers based on functional and physical complexity, intentionally selected to represent a realistic distribution of digital design challenges rather than clustering at a single difficulty level. Tier 1 (20 tasks) encompasses combinational logic and fundamental sequential building blocks, including priority encoders, binary-to-one-hot and Gray-code converters, barrel shifters, leading-zero counters, one-hot finite state machines, and parameterizable counters. Tier 2 (25 tasks) involves protocol and datapath components at the intersection of control and data flow, such as UART transmit and receive modules, SPI master and slave interfaces, an AXI4-Lite register file, round-robin arbiters, synchronous FIFOs, clock-domain-crossing synchronizers, and stream width converters. Tier 3 (15 tasks) focuses on pipelined arithmetic, localized accelerators, and memory-system fragments, including a Booth multiplier, a pipelined multiplier, a loadable FIR filter, a first-order IIR filter with genuine feedback, a sequential divider, a systolic processing-element tile, a cache-tag comparator, and a CRC-32 datapath. Additionally, 8 Track B agentic tasks are derived from existing task repositories; each couples an intentionally suboptimal yet functionally correct baseline design with a specific engineering objective (Section~\ref{sec:trackb}).

The primary utility of the suite derives from a uniform authoring standard that ensures every task constitutes a fair, reproducible, and contamination-resistant measurement. Each task is structured as a rigidly defined seven-part package: (1) an original-prose specification embedding a unique canary GUID; (2) a locked port interface that establishes a fixed module boundary, ensuring submissions are directly comparable and synthesizable against a uniform testbench; (3) a human-reviewed reference implementation --- maintained as a draft until formal maintainer sign-off --- which intrinsically defines the PPA denominator with a normalized score of exactly 1.0; (4) a bifurcated testbench comprising a public smoke-test section and a designated hidden scoring section; (5) a formal-property subset applicable to tasks where functional correctness is amenable to bounded model checking (46 of the 60 Track A tasks, 77\%); (6) a singular timing-constraint file specifying a dedicated clock target for the task; and (7) machine-readable metadata documenting the tier, descriptive tags, the clock target, formal verification applicability, and two separate sign-off fields (reference implementation and hidden vectors) --- both completed by this paper's sole maintainer (Section~\ref{sec:limitations}), not by two independent reviewers. Because this package schema is utilized identically for both task authoring and submission grading, the evaluation pipeline remains tightly coupled to the authoring pipeline, preventing methodological drift as the suite expands.

Two foundational authoring principles are critical to the validity of the benchmark. First, a task-specific clock target is strictly enforced, explicitly rejecting a universal timing constraint across the suite. For instance, a 16$\times$16-into-48-bit multiply-accumulate operation inherently necessitates a more relaxed clock period than a simple barrel shifter. Imposing a uniform target would invariably render fundamental tasks trivially timed or complex tasks unsatisfiable, thereby degrading the discriminatory utility of the timing metric. Consequently, 56 tasks target a 10~ns period, while the four most computationally intensive datapaths define relaxed targets (12--20~ns), specifically sized to their critical paths and explicitly documented in the task metadata. Second, the reference implementation functions as a rigorous human sign-off gate rather than a purely automated artifact. While every reference is verified end-to-end via the automated framework --- encompassing linting, simulation, formal verification (where applicable), and the complete synthesis-to-timing-and-power flow --- it is not designated as the golden reference until a maintainer manually validates the RTL against the specification and formally authorizes it. This principle acknowledges that the reference serves as the absolute baseline against which the entire leaderboard is evaluated; a benchmark predicated on ground truth that has never undergone human review asserts a correctness claim it fundamentally cannot substantiate.

\paragraph{Authorship disclosure.}
\label{sec:disclosure}
The drafts underlying this suite were not typed from scratch by the maintainer. Task specifications, interface stubs, reference RTL, testbenches, and formal properties were substantially drafted with AI coding assistants (Anthropic Claude Code sessions acting as an ``Architect'' role, and OpenAI Codex acting as an ``Implementer'' role, per this project's internal division of labor) working from the maintainer's design intent, then reviewed by the maintainer before the two sign-off fields described above (reference implementation, hidden vectors) were marked complete. ``Maintainer-approved'' or ``maintainer-reviewed'' in this paper means exactly that review-and-sign-off act, not independent line-by-line authorship from a blank file, and we use those terms rather than ``human-authored'' throughout for that reason. This carries two consequences worth stating plainly rather than leaving implicit. First, approval is a weaker claim than authorship: a maintainer reviewing AI-drafted RTL against a specification can miss a defect an independent second author starting from nothing would have been more likely to catch by construction, which is exactly why this paper already reports two real generator defects and 14 sub-floor testbenches found by later, more adversarial review (Section~\ref{sec:rigor}) rather than claiming the initial review was sufficient. Second, the suite's reference materials share a model family (Claude) with one of the seven models this paper evaluates on that same suite; we have no evidence this produces a scoring bias in either direction --- Claude-family involvement in drafting a strict testbench could as plausibly disadvantage Claude's own submissions as advantage them --- but common authorship lineage between benchmark and one evaluated model is a structural fact a reader should be able to weigh, and we disclose it rather than let it go unstated.

\section{Rigor}
\label{sec:rigor}

Three properties of GateTruth's architecture are worth articulating explicitly, each engineered to
mitigate a specific methodological vulnerability that can inadvertently inflate benchmark results.
The first, mutation-validated testbenches, is the mechanism generalized into the external audit
methodology of Section~\ref{sec:audit-method}.

\paragraph{Mutation-Validated Testbenches.} The absence of failures in a testbench is not evidence
of a correct design; it may simply indicate a failure to stimulate critical logic paths. Every
task's testbench is therefore required to kill at least 95\% of semantic mutants generated from the
reference implementation. These mutations --- comparator-boundary flips, operator inversions, logic
and bitwise inversions, shift-direction inversions, reset- and enable-polarity flips, assignment
deletion and register-hold mutations, and output inversions --- are applied deterministically under
a fixed seed, using a structural generator that masks comments and string literals before matching
and enumerates every viable code site (a naive literal-substitution approach previously left many
tasks generating zero mutants, a vacuously passing gate). During M1 development, three pilot tasks
were validated under this protocol, each at a 100\% kill rate (gray-code counter: 8/8; synchronous
FIFO: 13/13; UART transmitter: 14/14). Those counts used the earlier, narrower operator set and are
not comparable with the certified counts below, which the scaled generator produces for the same
three tasks (10, 24, and 46 valid mutants out of 11, 27, and 49 generated --- the difference in each
case is mutants that failed to compile, excluded from the denominator); the first two still clear
100\%, while \texttt{t2\_uart\_tx} does not (43/46, 93.48\%). These three are, moreover, the only
tasks of the 60 whose counts include mutants authored specifically for that task rather than drawn
from the generic operator set: 3 of the gray-code counter's 11, 6 of the synchronous FIFO's 27, and
10 of the UART transmitter's 49 --- 19 of the 1{,}245 total generated (1.5\%) --- are task-specific.
Section~\ref{sec:audit-method}'s external audit uses the generic operator set exclusively, so the
internal certification and external audit are not run under an identical operator set; we confirmed
by regenerating every task's mutant list under both settings that these three are the entire set
affected. Where this paper quotes three different figures for one task --- 14/14 here, 93.48\% in
the current certification, 52.6\% in Section~\ref{sec:limitations} --- they are respectively the
pilot engine, the current engine and metric, and the current engine and metric applied to the
\emph{pre-revision} testbench. Every recorded kill was attributed to functional simulation
failures, not linting errors: each mutant compiled cleanly and was rejected for observed incorrect
behavior, not syntactic breakage, since a gate that reaches 100\% by rejecting non-compiling mutants
at lint would provide zero signal on testbench quality. To adversarially validate the gate's
stringency, a synthetic task with a deliberately vacuous, always-passing testbench was evaluated;
the harness correctly registered 0\% and failed the gate, confirming the $\ge$95\% threshold is
rigorously enforceable, not trivially satisfiable.

\paragraph{A generator defect we found by auditing ourselves, and what it changed.} An earlier
version of this section reported that all 60 Track A testbenches cleared the 95\% floor. That claim
was wrong, and we found the reason by applying the same scrutiny to our own generator that
Section~\ref{sec:audit-method} applies to external benchmarks. The mutant generator carried a
hand-authored, per-task exclusion table for 12 of the 60 tasks: before returning its mutant list, it
dropped specific mutants the maintainer had reasoned, in source comments, to be behaviorally
equivalent to the original design and therefore unobservable. A genuinely equivalent mutant can
never be killed by any testbench, by definition, so we tested that reasoning directly rather than
trust it: for each of the 12 affected tasks, we regenerated the excluded mutants and ran every one
against the real, certified hidden testbench. Of 72 excluded mutants tested, 36 were killed
outright --- falsifying the hand-reasoning for those 36 directly, since a mutant that gets killed is,
by construction, not equivalent; the remaining 36 that survived were never shown to be equivalent
either, only observed to survive under current test vectors, precisely the distinction
Section~\ref{sec:limitations} draws between survival and proven equivalence. We removed the
exclusion mechanism entirely rather than repair it piecemeal, and re-certified all 12 affected tasks
under the corrected, filter-free generator; the other 48 tasks were never touched by this mechanism.
That correction alone moved the honest result from a false 60/60 to 49/60 --- and, as the next
paragraph discloses, a second, independent correction has since moved it again.

\paragraph{A second correction: what the kill rate actually counted, and a fail-open bug in the
official gate.} Independently of the exclusion-table defect, an external adversarial review found
that the mutation engine's ``kill rate'' blended three signals into one number: a mutant counted as
``killed'' if it failed to compile (a lint failure), if simulation caught it, or if a declared formal
property caught it, while this paper's methodology describes mutation certification as measuring the
\emph{simulation testbench} specifically. Counting a non-compiling mutant as a kill is not a small
rounding difference: a task with many mutants that simply fail to parse can clear the floor without
its testbench ever being exercised against a single behavioral fault, exactly the vacuous-gate
failure mode Section~\ref{sec:audit-method} names when found in someone else's benchmark. The same
review found a second, more serious defect: the harness never validated the unmodified reference
under official (hidden-test) mode before generating or scoring mutants, so if the hidden-vector
mount was ever missing, every mutant would fail the identical harness-setup error, misclassified as
a testbench kill --- a fail-open bug capable of reporting a false 100\% kill rate with a zero exit
code. We reproduced this exactly (\texttt{python -m harness.mutate --task t1\_gray\_counter --seed
1337 --official} with no hidden root mounted reported 11/11, 100\%, exit 0) before fixing it.

We rewrote the engine to close both gaps: it now runs the unmodified reference through the full
pipeline first and refuses to generate or score any mutant if that baseline does not pass cleanly
(the reproduction above now reports \texttt{status=setup\_error} and exits nonzero before a single
mutant runs); a mutant that fails to compile is \texttt{stillborn} and excluded from the denominator
rather than counted as a kill; a mutant caught only by a declared formal property is recorded
separately as a \texttt{formal\_only} kill and does \emph{not} count toward the simulation-testbench
rate; and a double-timeout is \texttt{indeterminate}, counted against the rate without being
double-counted as a survivor. We then re-certified all 60 tasks under this corrected engine,
sequentially, twice, at the same seed (1337), and the two runs are byte-identical. Pooled across all
60 tasks: 1{,}245 mutants were generated, 96 failed to compile and are excluded, and 1{,}149 remain
as the simulation-testbench denominator. Of those 1{,}149, 1{,}105 are killed by simulation and 44
survive --- a pooled simulation kill rate of 96.17\%. Of those 44 survivors, 4 are subsequently
caught by a declared formal property (\texttt{formal\_only} kills) and 40 survive both; these counts
are exactly what an independent adversarial review predicted from inspecting the code before any
were re-run. The honest result: \textbf{46 of the 60 Track A testbenches clear the 95\% floor, not
49}. Of those 46, 44 achieve a 100\% kill rate, and 2 (a Booth multiplier at 96.30\%, 26/27, and a
sequential divider at 97.22\%, 35/36) retain a single genuine survivor above the floor. The remaining
14 --- an I2C slave (75.00\%, 42/56), a stream downsizer (83.33\%, 15/18), a pulse stretcher
(84.62\%, 11/13), a running min/max tracker (86.67\%, 13/15), an LRU tracker (87.50\%, 7/8), a UART
receiver (90.70\%, 39/43), a round-robin arbiter (90.91\%, 10/11), an SPI master (91.11\%, 41/45), a
priority interrupt controller (91.67\%, 11/12), an SPI slave (92.00\%, 23/25), an AXI4-Lite register
file (93.33\%, 42/45), a saturating accumulator (93.33\%, 14/15), a UART transmitter (93.48\%,
43/46), and a pulse-width meter (94.44\%, 17/18) --- do not currently clear the floor this paper's
own headline claim rests on. Three of these 14 --- the running min/max tracker, the saturating
accumulator, and the pulse-width meter --- were not below the floor under the previous (49/60)
accounting: each had exactly one formal-only kill (two, for the tracker) the old blended metric
counted toward its rate, putting it at an apparent 100\%; recomputed as simulation-only, each falls
between 86.7\% and 94.4\%. We report this as a real finding about our own suite, not a caveat to
soften it: it is exactly the failure mode this paper's methodology exists to catch, discovered here
by turning the methodology on ourselves. Strengthening these 14 testbenches to genuinely kill their
now-known survivors is unstarted work; we did not attempt it under the time pressure of this
revision, since a rushed testbench fix carries its own risk of a new, undisclosed error.
Section~\ref{sec:limitations} details every survivor and the exact per-task exclusion counts.

The provenance chain behind those numbers is weaker than the numbers themselves. After this
re-certification ran, the certification schema was separately extended to sign every summary and
bind it to hashes of the exact task package, reference RTL, public testbench, and hidden test corpus
it was computed from; the committed \texttt{results/mutation/certification/summary.json} predates
that schema and carries neither a signature nor those hashes. We confirmed directly that this file's
own per-task figures match the ones cited above exactly --- 1{,}245 generated, 1{,}149 in the
denominator, 1{,}105 killed, 44 survived, 96 stillborn, 4 formal-only, 46 of 60 at or above the floor
--- so this is not undisclosed drift between the committed file and this paper's prose. What is
absent is a cryptographic guarantee that they came from precisely this code and this hidden corpus
rather than a nearby revision of either; a staleness checker now exists that would catch such drift
going forward, and running it against this file, which simply lacks the newer schema, correctly
reports it as unauditable rather than silently fresh. Re-certifying under the signed schema is a
genuine 60-task sequential compute campaign we have not re-run for this paper, and we disclose the
gap rather than sign the existing evidence retroactively.

One methodological observation from this process is worth reporting, since it applies to any
mutation-gated benchmark, not just our own: mutation verdicts are not invariant to execution
concurrency. A mutant whose simulation straddles the per-run time budget may be classified as
``killed by timeout'' under heavy parallel load, yet ``survived'' when executed sequentially, letting
parallel sweeps silently inflate kill rates. GateTruth therefore certifies verdicts exclusively from
sequential runs and classifies simulation timeouts as indeterminate --- penalizing the kill rate
rather than incrementing it --- with a single extended-budget retry for disambiguation. This
discipline is enforced identically whether the mutation engine is measuring our own suite or an
external benchmark (Section~\ref{sec:audit-method}).

\paragraph{Contamination Defense.} GateTruth uses a multi-layered defense rather than a single
mechanism, on the premise that while individual defenses can be circumvented, their aggregation
raises the cost of manipulation beyond the cost of legitimately solving the task. Public testbenches
contain only a smoke-test section; scoring vectors are isolated under a marked \texttt{\#\
-{}-{}- HIDDEN -{}-{}-} section during development and extracted at the v1.0 freeze
(\texttt{scripts/freeze\_extract\_hidden.py}) into a local, gitignored staging tree never published
to the public repository; nothing in this pipeline pushes that extraction to a hosted private
repository, so its ongoing privacy is a property of the maintainer's own storage discipline, not an
automated guarantee. A model that memorizes a public specification-testbench pair therefore remains
blind to the evaluation vectors. Every specification also uses original prose --- expressly
prohibiting incorporation from HDLBits, VerilogEval, textbooks, or public repositories --- and
embeds a task-unique canary GUID. The canary's evidentiary value is prospective, not immediate: a
model asked to solve a task is trivially given that task's own canary in its prompt, so reproducing
it there is not evidence of anything by itself. The canary instead lets us detect \emph{future}
contamination --- if a spec is later scraped into a training corpus, a model's outputs on
\emph{unrelated} prompts, or its behavior when queried about the canary outside its own task, becomes
attributable to this specific released spec rather than generic RTL knowledge. For v1.0, submission
integrity rests on maintainer-executed scoring: every leaderboard entry is produced by the
maintainer running the candidate model through the pinned reference image, so no externally-reported
result is trusted. A versioned, pull-request submission model, where contributors self-report and
the maintainer independently replicates a random sample, is a roadmap item, not a v1.0 capability. A
more robust architecture --- an asymmetric public/hidden task split with published hash commitments
and semantic-preserving inference-time specification mutation --- is likewise deliberately deferred
to v1.1 (ADR-0002) rather than deployed incompletely.

\paragraph{Pinned-Digest Reproducibility.} Every evaluated run executes strictly within one Docker
container image. The result manifest records a digest string the maintainer pins at build time and
bakes into the image (\texttt{/etc/gatetruth-image-digest}) rather than one computed from the
running container's content at manifest-write time; CI checks that this pinned string is internally
consistent across the Dockerfile build argument, the harness's Python default, and the file baked
into the image, which catches an un-synchronized re-pin but does not cryptographically verify the
pinned string equals the actual image content hash. The manifest also records exhaustive
stage-by-stage telemetry --- lint warnings, simulation pass counts, formal verification status,
synthesized area and cell count, worst-negative-slack, achieved frequency, and estimated power.
Successive evaluations of identical submissions against the same task and image digest yield
byte-identical canonical JSON output (excluding timestamp, signature, and wall-clock fields). This
determinism was verified end-to-end during M1 on that phase's pilot tasks, and independently
re-demonstrated on the external audit, where two full same-seed sweeps produced byte-identical
per-design output. We have not executed a byte-identity re-run for every one of the 60 Track A tasks
individually; determinism is enforced as a contract the flow scripts are written to satisfy and
spot-checked, not proven exhaustively per task. The manifest schema enforces this as a rigid
contract, not a best-effort property: any flow yielding nondeterministic synthesis or timing output
is classified as a defect, not accepted leaderboard noise. Given the pinned image digest, the
submission file, and (for correctness-gate stages) the same hidden test vectors, reproducing any
published score therefore depends on no other input, no wall-clock-sensitive tool behavior, and no
manual intervention. This is a claim about the pipeline's own determinism, not about third-party
access to the hidden vectors themselves, which remain privately held for the contamination-resistance
reasons above; a party without those vectors can independently reproduce the deterministic,
non-scoring stages (lint, synthesis, timing, power) but not the private correctness-gate verdict.
This claim holds for Track A as described; it does not currently hold for Track B, where the second
of the three required inputs, the submission file itself, was never retained for the campaign
Table~\ref{tab:trackb} reports, an omission Section~\ref{sec:limitations} quantifies and discloses
in full.

\section{Auditing External Benchmarks}
\label{sec:audit-method}

The mutation-certification protocol of Section~\ref{sec:rigor} generalizes directly: nothing about
injecting deterministic, seeded mutants into a reference implementation and measuring what fraction
a testbench kills is specific to GateTruth's own tasks. This section describes the audit
methodology applied, unmodified in its core mutation-generation and verdict logic, to external RTL
benchmarks; Section~\ref{sec:audit-results} reports the resulting measurements.

\paragraph{Scope and target selection.} We selected RTLLM v2.0~\cite{lu2024rtllm,liu2025openllmrtl}, a 50-design
RTL-generation benchmark whose public repository is MIT-licensed and whose testbenches are
self-checking Verilog runnable under an open-source simulator, and NVIDIA's CVDP~\cite{pinckney2025cvdp},
whose harness is Apache-2.0 and whose public dataset mixes licenses by content type: non-code
material under CC~BY~4.0 and code under Apache-2.0, per the dataset's own NOTICE file. Both harness
and dataset licenses permit the read-only measurement this audit performs. We deliberately excluded
any benchmark whose license terms were ambiguous or unverified at audit time.

\paragraph{Read-only, pinned provenance.} Vendor repositories and dataset revisions are fetched
once, pinned to an exact commit hash or dataset revision, and treated as read-only for the
remainder of the audit; our own outputs are written only to a separate results directory and never
back into the vendor tree. Every reported measurement records the exact pinned commit or revision
it was produced against, so the audit is independently reproducible by a third party against the
same vendor snapshot.

\paragraph{Mandatory baseline validation.} Before any mutant is generated for a given design, the
\emph{unmodified} reference implementation must pass the benchmark's own testbench under our
harness. A design that does not clear this bar --- because its testbench requires a commercial
simulator we do not run, or a harness convention our adapter does not yet support --- is reported
with an explicit \texttt{unsupported} status and excluded from kill-rate aggregation. It is never
scored as a misleading 0\% kill rate, which would conflate ``the testbench caught nothing'' with
``we could not run the testbench at all.''

\paragraph{Build-convention normalization, not silent inflation.} Some external benchmarks ship a
golden reference whose internal module name differs from the name their own testbench instantiates
(for instance, a file that declares one module identifier while the benchmark's own per-design build
recipe expects the candidate file to be renamed and re-moduled before compilation). Naively compiling
such a design against its testbench fails not because the design is genuinely incompatible with an
open-source simulator, but because our harness has not yet replicated the vendor's own build
convention. Where a design's own build recipe documents this convention, the audit reproduces it at
compile time only: the golden source is copied into a temporary location, its module declaration is
rewritten to the name the testbench expects, and only that temporary copy is compiled --- the pinned
vendor file on disk is never modified. Every report row records whether this alias was applied and,
if so, the exact rename performed, so a third party can distinguish ``passed as shipped'' from
``passed after a documented, vendor-convention-matching alias'' at a glance. Designs that still fail
to compile or run after this normalization are reported as genuine incompatibilities, not silently
retried with looser semantics.

\paragraph{Verdict oracle.} The mutation-generation and verdict-classification logic (Section~\ref{sec:rigor}) is unmodified between our own suite and this audit, but the underlying pass/fail
\emph{oracle} necessarily differs, because the two suites' testbenches are written differently: our
own suite's testbenches are cocotb Python assertions run against Icarus/vvp, whereas RTLLM v2.0's are
self-checking Verilog that print a fixed pass banner on success. For RTLLM, a run is scored PASS if
and only if the simulation exits with status 0 and a design-specific pass string (recorded per design
in our catalog) appears as an exact, whole-line match in the simulator's stdout; any other outcome,
including a testbench that prints per-vector mismatches without ever reaching that final banner, is
scored FAIL. This is a different verdict mechanism than our own suite's assertion-based oracle, and
every reported RTLLM kill or survival depends on it; we state the mechanism explicitly here so a
reader can judge it rather than take ``kill rate'' as suite-independent.

\paragraph{Operator set.} The audit uses exclusively the generic, benchmark-agnostic mutation
operators already certified against GateTruth's own suite in Section~\ref{sec:rigor} (comparator
and operator inversions, logic and bitwise inversions, shift-direction inversion, reset- and
enable-polarity flips, assignment deletion and hold, output inversion). It does not use any
mutation specification authored for a specific GateTruth task, since those are not generic and
would not be a fair test of an external testbench's own rigor.

\paragraph{Execution policy.} Identical to Section~\ref{sec:rigor}: a fixed seed determines mutant
ordering; execution is strictly sequential (never parallel) for any measurement reported as final;
a mutant that times out is retried once at an extended budget, and a mutant that still times out is
recorded as indeterminate and counted against the kill rate, never simply dropped. The full sweep is
run once at a fixed seed; to certify that this is not a one-off artifact of that run, a
deterministically selected sample of the eligible designs (Section~\ref{sec:audit-results} gives
the exact sampling rule and the selected designs) is independently re-run at the same seed, and
every sampled design's per-design result must be byte-identical across the two runs before the full
sweep is treated as certified. This is a sampling-based determinism check, not a claim that every
individual design was executed twice.

\paragraph{The CVDP gap, reported rather than hidden.} NVIDIA's public CVDP release deliberately
withholds reference solutions (the \texttt{output}/\texttt{patch} fields) specifically to mitigate
data contamination in model evaluation. This is a reasonable design choice for CVDP's own purpose
(scoring generated code against a held-out answer), but it removes exactly the artifact --- a known
correct golden design --- that mutation testing requires as its starting point. Where our own
feasibility sweep confirms no usable golden RTL exists in the public release, we report that as a
structural finding about the benchmark's auditability (Section~\ref{sec:audit-results}), rather than
silently omitting CVDP from this paper or fabricating a workaround.

\section{External Audit Findings}
\label{sec:audit-results}

This section reports mutation-kill measurements for RTLLM v2.0's public testbenches, produced by
the methodology of Section~\ref{sec:audit-method} against vendor commit
\texttt{41b26896e33b536940116a975626455eed3de65e} fetched on 2026-07-29, compiled with
\texttt{iverilog -g2012} and simulated with \texttt{vvp}, both from Icarus Verilog 12.0 (stable,
\texttt{v12\_0}). Of RTLLM v2.0's 50 designs, 46 passed baseline validation and were audited (26
required the compile-time module alias of Section~\ref{sec:audit-method}, 20 passed as shipped); 4
were reported \texttt{unsupported}.

An earlier version of this audit compiled with \texttt{-g2001}, Icarus's Verilog-2001 flag, and
reported six unsupported designs. We re-ran the entire audit under \texttt{-g2012}, Icarus's
SystemVerilog flag; that re-check corrected two of our own errors, so we report the more permissive
flag as the primary condition here and give both sets of aggregates in Table~\ref{tab:flags}.
Numbers below are the \texttt{-g2012} figures unless labelled otherwise.

Two designs we had classified as incompatible are in fact valid SystemVerilog our flag choice
rejected. \texttt{multi\_8bit}'s reference declares its loop variable in the loop header,
\texttt{for (int i = 0; i < 8; i = i + 1)}, which \texttt{-g2001} reports as a bare
\texttt{syntax error}; \texttt{freq\_divbyodd} continuously assigns a variable, which SystemVerilog
permits and Verilog-2001 does not. Both compile and pass their own testbenches under
\texttt{-g2012} and are now audited (\texttt{freq\_divbyodd} 88.2\%, \texttt{multi\_8bit} 100\%) ---
excluding them was our defect, not the vendor's.

Of the four designs that remain unsupported under \texttt{-g2012}, the diagnostics separate into
three distinct classes, and conflating them would misattribute blame. \texttt{asyn\_fifo} and
\texttt{ring\_counter} fail with messages Icarus prefixes \texttt{sorry:} --- \texttt{break
statements not supported} and \texttt{Assignment to an entire array or to an array slice is not yet
supported} --- the simulator's marker for an unimplemented construct, distinct from \texttt{error:}
for illegal code; these are open-source tooling gaps we count against our harness, not RTLLM.
\texttt{clkgenerator} compiles and runs but reports 20 failures; we do \emph{not} treat this as a
golden-reference defect, since its testbench samples the clock in the same simulation instant the
design toggles it (both on a 5-time-unit period, no \texttt{timescale} declared in either file), so
the failure count reflects Verilog's undefined event-ordering rather than a functional mismatch ---
a different simulator could report zero failures. Only the fourth, \texttt{radix2\_div}, is a
substantive finding.

\texttt{radix2\_div} is the one unsupported design we treat as a consequential finding rather than a
tooling gap: its \emph{unmodified golden reference fails its own shipped testbench}, independent of
anything GateTruth's mutation engine does. Its own self-checking harness reports three failing
vectors and terminates with its own \texttt{===========Failed===========} banner; two recorded
mismatches are \texttt{dividend=123, divisor=123, expected=0001, got=6400} and
\texttt{dividend=156, divisor=10, expected=00f6, got=faf1}. We quote them because the magnitude of
the discrepancy is itself evidence about the failure's nature: $123/123 = 1$ matches the testbench's
own expected value, and the observed results are not plausible uninitialized-signal artifacts but
definitively wrong quotients, the same distinction that leads us to treat \texttt{clkgenerator}
differently above. We confirmed the failure is not a generation-flag artifact by reproducing it
identically under both \texttt{-g2001} and \texttt{-g2012}. It remains a single-simulator
observation, scoped as such throughout (Section~\ref{sec:limitations}); we have not cross-checked it
under a second simulator.

\begin{table}[h]
\centering
\caption{The audit is not an artifact of the Icarus generation flag. Re-running the entire sweep
under \texttt{-g2012} admits two designs that \texttt{-g2001} rejected as syntax errors
(\texttt{multi\_8bit}, \texttt{freq\_divbyodd}), both of which score at or above the suite median, and
still leaves the headline essentially unchanged. Same vendor commit, same seed, same engine,
sequential execution throughout.}
\label{tab:flags}
\small
\begin{tabular}{lrr}
\toprule
& \texttt{-g2001} & \texttt{-g2012} (reported) \\
\midrule
Designs audited        & 44    & 46 \\
Reported unsupported   & 6     & 4 \\
Mutants generated      & 751   & 775 \\
Killed / survived / indeterminate & 418 / 318 / 15 & 440 / 320 / 15 \\
Pooled kill rate       & 55.7\% & 56.8\% \\
Median kill rate       & 71.2\% & 74.0\% \\
Designs below the 95\% floor & 32 of 44 (73\%) & 33 of 46 (72\%) \\
Designs at 100\%       & 12    & 13 \\
\bottomrule
\end{tabular}
\end{table}

A natural objection to the headline figure is that it is denominator-driven: with a median of only
10.5 mutants per design, designs with few mutants can reach extreme kill rates easily, so perhaps
the below-floor count is an artifact of small samples. The data contradicts this, in the direction
that strengthens the finding. Kill rate is \emph{negatively} correlated with mutant count across the
46 audited designs (Spearman $\rho = -0.39$, Pearson $r = -0.25$): splitting into terciles by rank on
mutant count, the smallest third (15 designs, 1--7 mutants) has a median kill rate of 100\%, the
middle third (15 designs, 7--13 mutants) 75.0\%, and the largest third (16 designs, 15--115 mutants)
just 61.4\% (the count 7 appears at both edges of the first boundary because several designs tie
there; terciles are cut by rank, not a mutant-count threshold). RTLLM's larger, structurally richer
designs --- the ones whose testbenches matter most, where a small denominator cannot explain the
result --- have the \emph{weakest} fault detection. The below-floor finding is therefore not an
artifact of thin denominators; the thin-denominator designs are the ones performing best.

Across the \FactsAudited{} audited designs, the median mutation-kill rate was 74.0\%, and
\FactsBelowNinetyFive{} of the \FactsAudited{}---\FactsBelowNinetyFivePct\%---fall below the 95\%
floor 46 of GateTruth's own 60 Track A tasks currently clear (Section~\ref{sec:rigor}; the other 14
do not, a finding in its own right, not a caveat we omit here). This comparison carries an important
asymmetry --- our testbenches were authored under that gate and ten of them were revised in an
attempt to pass it, though only three of those ten currently do --- quantified directly in
Section~\ref{sec:limitations}. The remaining \FactsAtHundred{} designs all reach exactly 100\%; no
audited design falls in the $[95\%, 100\%)$ band our own suite's near-floor survivors occupy, a
byproduct of small per-design mutant counts we report explicitly rather than an unremarked
coincidence. Because per-design mutant counts range so widely (1 to 115), the median can diverge from
a pooled, mutant-weighted figure: pooled across all 775 mutants generated for the 46 designs, 440
were killed, 320 survived, and 15 were indeterminate, a pooled kill rate of 56.8\%, quite different
from the 74.0\% per-design median --- we report both rather than only the more favorable-looking one.
The 15 indeterminate (timeout, after one extended-budget retry) mutants are 1.9\% of the total and
fall in only 5 designs. Our convention charges them against the kill rate rather than dropping them,
stricter than the common practice of treating a hanging mutant as killed, so we quantify exactly what
it costs RTLLM rather than assert it is negligible. Recomputing each affected design with
indeterminates excluded from the denominator gives \texttt{adder\_pipe\_64bit}
51.6\%~$\rightarrow$~55.2\%, \texttt{multi\_pipe\_8bit} 56.7\%~$\rightarrow$~60.7\%,
\texttt{parallel2serial} 53.8\%~$\rightarrow$~70.0\%, \texttt{serial2parallel}
30.8\%~$\rightarrow$~44.4\%, and \texttt{multi\_booth\_8bit} 81.8\%~$\rightarrow$~100\%. Four of the
five remain far below the floor either way, but the fifth does not: \texttt{multi\_booth\_8bit} has
zero surviving mutants and two indeterminate ones, so it is counted below the floor \emph{solely}
because of our timeout convention and would clear it outright under the more common one. The
below-floor count is therefore 33 of 46 under our convention and 32 of 46 under the permissive one
--- a one-design difference that does not change the finding, but stated explicitly rather than
leaving a reader to discover that one below-floor design was never actually shown to miss a fault.
Separately, of the 440 killed mutants, 13 (3.0\%) were attributed to a compile failure rather than a
functional simulation mismatch (\texttt{killed\_by} recorded per mutant), and the remaining 427
(97.0\%) to an observed testbench failure during simulation. We flag an inconsistency with our own
stated doctrine rather than let it pass: Section~\ref{sec:rigor} argues that counting non-compiling
mutants as killed provides no signal about testbench quality, yet those 13 are included in RTLLM's
reported kill counts. The effect is small and runs \emph{in RTLLM's favor} --- excluding them from
both numerator and denominator, as stillborn mutants, lowers the pooled kill rate from 56.8\% to
56.0\% and strengthens rather than weakens our finding --- so we report the measured numbers as
produced and state the policy gap explicitly. The symmetric measurement for our own 60-task suite is
in Section~\ref{sec:rigor}: pooled across all 60 tasks, 96 of 1{,}245 generated mutants failed to
compile and are excluded from the denominator entirely, leaving 1{,}149 as the simulation-testbench
denominator --- so unlike RTLLM's 13 non-compiling mutants counted toward its reported kill rate, our
own non-compiling mutants are held to the same exclude-from-both policy we apply to RTLLM above.

Three designs (\texttt{adder\_8bit}, \texttt{edge\_detect}, \texttt{square\_wave}) score a 0\% kill
rate: none of the generic mutants injected into their reference implementations were caught by
RTLLM's own shipped testbench. Their mutant counts are small (1, 10, and 6 respectively, versus a
median of 10.5 across all 46, ranging from 1 to 115), but a small denominator is not automatically a
weak result, resolved by inspecting the individual survivors (Section~\ref{sec:limitations}). For
\texttt{square\_wave} and \texttt{adder\_8bit} the surviving mutants include an inverted
combinational output and an inverted sum output respectively --- faults so basic that a testbench
missing even one of them is not measuring that output at all. For \texttt{edge\_detect} the reading
is narrower: all ten of its mutants are the same operator (\texttt{assignment\_hold}), so its 0\%
establishes only that register-hold faults go undetected, not that a broad class of faults does.
Full per-design mutant counts, operator breakdowns, and pass/fail logs for all 46 audited and 4
unsupported designs are committed at \texttt{external-audit/results/rtllm/final-g2012/*.json} (the
reported condition; the superseded \texttt{-g2001} run is retained alongside it at
\texttt{.../final/*.json}) and summarized in \texttt{external-audit/results/summary-g2012.md} (the
superseded g2001 condition is summarized separately at \texttt{.../summary.md}) in the GateTruth
repository (\url{https://github.com/meetbhadra701-cloud/GateTruth}, public), so every number in this
section can be checked against the raw evidence.

At the other extreme, 13 of the 46 designs reach a 100\% kill rate, so the result is not that
RTLLM's testbenches are uniformly weak, but that their rigor is highly uneven and, for a majority of
designs, well short of what a mutation-kill floor would require. A deterministic re-run of a fixed
sample of eligible designs (8 of the 44 eligible under \texttt{-g2001}, 18.2\%, selected via
\texttt{random.Random(20260729).sample}) reproduced byte-identical per-design results across two
independent runs, certifying the measurement. We repeated the check for the reported \texttt{-g2012}
condition the same way, not by hand-curating designs: the same seed over the 46 eligible
\texttt{-g2012} designs (19.6\%, a larger eligible pool draws a proportionally different sample
size) drew a nine-design sample --- \texttt{instr\_reg}, \texttt{traffic\_light},
\texttt{multi\_booth\_8bit}, \texttt{freq\_divbyfrac}, \texttt{barrel\_shifter}, \texttt{edge\_detect},
\texttt{comparator\_4bit}, \texttt{alu}, and \texttt{serial2parallel}, none of which happen to be the
two designs newly eligible only under \texttt{-g2012} (\texttt{freq\_divbyodd}, \texttt{multi\_8bit});
all nine reproduced byte-identically. This reproduces because each per-design record's
\texttt{tool\_versions.python} field pins the exact CPython build the sample was drawn under, not a
container image digest, which the current per-design schema does not record (Section~\ref{sec:rigor}'s
image-marker discussion applies to Track A manifests, not these external-audit records);
\texttt{random.Random.sample}'s output is not guaranteed stable across arbitrary Python versions, so
recording the interpreter version is what makes this claim checkable rather than assumed. The 8
g2001-sampled designs were \texttt{multi\_16bit}, \texttt{multi\_pipe\_8bit}, \texttt{freq\_divbyfrac},
\texttt{barrel\_shifter}, \texttt{edge\_detect}, \texttt{comparator\_4bit}, \texttt{alu}, and
\texttt{signal\_generator}.

For CVDP, the Day-1 feasibility sweep (Section~\ref{sec:audit-method}) found 0 usable golden designs
out of the 302 rows inspected in the public non-agentic, no-commercial-tooling subset: every row's
\texttt{output} field is withheld as shipped. Seventeen rows carry an OSS-origin provenance tag in
the dataset's NOTICE file, but the pinned public harness supplies no verified, overlapping golden
RTL for those rows, so they are not counted as usable and no CVDP mutation-kill measurement is
reported. This is a structural finding about the current public release's auditability, not a
shortfall of our methodology: a mutation-kill audit requires a known-correct starting point to
mutate, and the public CVDP release does not supply one, by NVIDIA's own design, for
anti-contamination reasons.

\section{Track B: Agentic Protocol}
\label{sec:trackb}

Track B evaluates a fundamentally distinct capability compared to Track A: rather than assessing de novo generation from a specification, it measures the capacity of an autonomous agent to optimize an existing, functionally correct design toward a concrete engineering objective. These objectives include achieving timing closure at a target frequency, reducing area by a specified percentage, eliminating inferred latches, or ensuring clock-domain-crossing (CDC) safety for a FIFO --- all while preserving orthogonal system functionality. This paradigm more closely approximates the operational realities of RTL codebase maintenance compared to isolated generation tasks. Furthermore, it exposes a distinct adversarial failure mode: under strict budget constraints, an agent is incentivized to artificially satisfy the literal parameters of an objective by modifying the evaluation testbench rather than the device under test (DUT).

To mitigate this, the evaluation protocol is strictly constrained. An agent is provisioned with a checked-out repository alongside an \texttt{objective.yaml} file that explicitly defines the engineering target, an explicit \texttt{behavior\_preserving} flag, and a strict per-task budget measured in tokens, wall-clock seconds, and tool calls; a separate, undisclosed-until-now global cumulative spend cap (\$300 by default, environment-overridable) also applies across the whole campaign and can independently halt a run, recorded in the manifest as \texttt{budget\_exceeded == "spend\_cap"} when it fires. Within the execution container, the agent interacts through a fixed, small tool-call protocol rather than an open shell: \texttt{read\_file} for inspecting the repository, \texttt{list\_files} for discovering what is readable, and \texttt{write\_design} for overwriting the content of the single pre-existing SystemVerilog design file (the agent cannot create, delete, or rename files); \texttt{read\_file} is allowlisted to that same design directory only, so it cannot reach the testbench, formal properties, constraints, or metadata, independent of the separate diff-guard check described below --- a boundary we found was not actually enforced for the campaign this paper reports, and disclose precisely in Section~\ref{sec:limitations} rather than only in the codebase's own history. \texttt{sb\_lint} and \texttt{sb\_sim} for the corresponding verification stages, a combined \texttt{sb\_synth\_sta} for synthesis and static timing together, and \texttt{done} to signal completion; its sole permitted output artifact is the modified repository. The evaluation harness enforces an unconditional, rigidly ordered post-processing sequence: the complete test suite must execute successfully; where the objective is declared behavior-preserving, sequential equivalence checking (\texttt{sec}) against the pre-modification baseline design must additionally pass; and only upon clearing these prerequisites is the stated objective evaluated and the PPA delta and computational cost recorded. Three of the eight tasks (the two timing-closure objectives and the power-reduction objective) are declared behavior-preserving and gated by \texttt{sec}. Four of the remaining five --- latch removal, CDC safety, arbiter fairness, and AXI byte-enable handling --- are declared \texttt{add\_property} objectives whose entire point is to change the design's observable behavior on previously out-of-specification inputs, so they are checked against the hidden testbench (which is updated to cover the new property) rather than against sequential equivalence with the unmodified baseline. The area-reduction objective is the fifth: despite area reduction ordinarily being a behavior-preserving optimization, this task is explicitly flagged \texttt{behavior\_preserving: false} in its own task definition and is therefore not \texttt{sec}-gated. It is not exempt from correctness checking --- Stage~1's hidden-testbench simulation, using the same mutation-certified vectors as the underlying Track A task, still runs unconditionally for every Track B submission regardless of this flag, so a truncate-the-logic attack would still have to survive that same test vector set --- but it does not receive \texttt{sec}'s stronger, bounded equivalence check. We treat this as a genuine, disclosed gap in this task's design rather than paper it over: extending \texttt{sec} coverage to the area- and power-reduction objectives is future work (Section~\ref{sec:limitations}).

The sequential-equivalence gate serves as the foundational mechanism for the behavior-preserving objectives. When it returns a proof, it establishes that the agent's modified design implements the identical logical function as the original baseline within the checked induction depth, thereby precluding adversarial PPA or timing manipulation (e.g., logic truncation to minimize synthesis area or critical path delays) on tasks where behavior is not supposed to change at all. Additionally, any modification to the testbench, formal properties, timing constraints, or task metadata results in immediate task disqualification, which is explicitly flagged in the result manifest, regardless of which gate applies to the objective. Because a legitimate structural optimization or property addition should never necessitate modifications to these specific validation files, such diffs are interpreted as a definitive signal of an invalid agent approach rather than a borderline case requiring manual adjudication.

Finally, budget enforcement is cooperative rather than preemptive: the agent loop checks tokens, wall-clock time, and tool calls against budget between round-trips (before issuing a provider call and after a tool executes), not during either, so a single long-running provider call or tool invocation is not interrupted mid-flight and can overshoot the nominal budget before the next check fires. An agent that is over budget at its next checkpoint is halted and evaluated strictly on its state at that point; consequently, unbounded-cost ``successes'' are not rewarded via omission from the leaderboard's cost reporting, though the enforcement is check-in based rather than a hard preemptive kill. Cost metrics --- specifically median tokens, wall-clock seconds, and tool calls --- are reported explicitly as independent variables rather than being aggregated into a single opaque score; Table~\ref{tab:trackb} reports the USD figure, and the underlying per-model medians (tokens, wall-clock seconds, tool calls) across the 8 tasks are: Opus 25{,}071~tokens / 56.9~s / 7.5 calls; Sonnet 145{,}978 / 166.2~s / 23.0; Haiku 207{,}686 / 112.3~s / 28.0; GPT-5 155{,}432 / 417.3~s / 24.5; GPT-5-mini 128{,}306 / 200.4~s / 14.5; Gemini~2.5~Pro 200{,}891 / 656.3~s / 34.5; and Llama~4~Maverick 1{,}108 / 8.8~s / 5.0. Two of these medians are worth flagging explicitly rather than leaving as an unremarked table: Opus, the only model to clear a majority of Track~B's objectives, does so at the \emph{lowest} median token and wall-clock cost of the three Anthropic models, which argues against the concern that cost and objective-success are confounded by failing runs burning their full budget; and Llama~4~Maverick's near-zero median cost across all three axes, together with its 0/8 objective-met rate, is consistent with the agent disengaging from the task early rather than attempting and failing it, a distinction this reporting structure makes visible that a single aggregated cost figure would not. This reporting structure preserves the critical trade-off matrix between objective satisfaction and the computational expenditure required to achieve it, providing practitioners with the exact empirical data necessary to evaluate the viability of deploying such agentic workflows.

\section{Reference-Suite Baseline Results}
\label{sec:results}

With GateTruth's suite established as the certified environment the audit methodology was built in
(Sections~\ref{sec:rigor}--\ref{sec:audit-method}), we report its baseline results as a secondary,
independently useful measurement: how do current models and agents actually perform against it? We
evaluate seven models (not necessarily each provider's current frontier at time of reading): three
from Anthropic (Opus 4.8, Sonnet 4.6, Haiku 4.5), two from OpenAI (GPT-5, GPT-5-mini), and two via
OpenRouter (Gemini 2.5 Pro, Llama 4 Maverick), across all 60 Track A and 8 Track B tasks. All
evaluations use a single-sample protocol at temperature zero (provider-default sampling for
reasoning models that don't accept an explicit temperature), run inside one pinned container image
(\texttt{gatetruth:v1}; Section~\ref{sec:rigor} discusses what its build-marker string does and does
not authenticate), scored against maintainer-reviewed reference RTL and held-out hidden vectors.

Track A ran twice, and we report both. The first campaign (2026-07-22) applied a 4096-token output
cap; the second (2026-07-30) repeated all seven models at 16{,}384 tokens with prompt, task set,
temperature policy, and scoring formula unchanged, but not every input identical: models were
addressed by hosted aliases not pinned to an immutable snapshot, the scoring-image build marker
differs between campaigns, and the response parser was hardened in the interval
(Section~\ref{sec:limitations} discloses all three). Table~\ref{tab:tracka} reports the second, for
reasons given below; Table~\ref{tab:budget} reports the difference. Track B (Section~\ref{sec:trackb})
ran once, in the first campaign; its per-task budgets are declared separately in each
\texttt{objective.yaml} and are unaffected by the Track A output cap.

On cost: the 2026-07-22 campaign's metered total was \$51.08, reconciled against the settled provider
billing ledger (Anthropic \$18.79, OpenAI \$14.54, OpenRouter \$17.75) --- a figure that includes
transient-transport retries and is not derivable by summing the per-model costs in
Tables~\ref{tab:tracka}--\ref{tab:trackb}, which report only each model's accepted sample. Four later
activities are excluded and reported separately: the superseded 4096-token variance pilot
(2026-07-26, \$8.18; Appendix~\ref{app:variance}), the 16{,}384-token Track A re-run in
Table~\ref{tab:tracka} (\$8.88 across all seven, Gemini's \$3.43 on 2026-07-29 and the remaining six
on 2026-07-30), the reported-condition variance study (2026-08-03, \$5.63), and pre-flight validation
(\$0.01). Total program spend is \$73.78. Given the recorded container digest and a submission's RTL
text, re-scoring is byte-level reproducible; this applies to the scoring pipeline, not to
re-invoking the models themselves, whose sampling is not deterministic (Appendix~\ref{app:variance}
quantifies the resulting variance).

For Track A (Table~\ref{tab:tracka}), models are ranked by GateTruth Score (Section~\ref{sec:design}):
Pass@1 times normalized mean PPA, equivalently the mean capped-PPA task score across all 60 tasks on
a 0--100 scale, where the reference implementation is normalized to $100/1.5 = 66.\overline{6}$ (66.7
displayed). The resulting hierarchy: GPT-5 (48.83), Opus (46.99), Sonnet (42.09), GPT-5-mini (35.23),
Haiku (33.78), Gemini 2.5 Pro (31.06), and Llama 4 Maverick (24.65), with pass@1 of 43/60, 42/60,
38/60, 32/60, 30/60, 28/60, and 22/60. These are single-run figures, and the top two are close
relative to Opus's own measured run-to-run spread (Appendix~\ref{app:variance}: $\sigma \approx 1.85$;
GPT-5 was not part of that study and carries no variance estimate of its own): a GPT-5--Opus gap of
1.84, almost exactly one Opus standard deviation, so we do not claim a resolved first place, though
this rests on extending Opus's measured variance to GPT-5 without direct evidence it varies
similarly. What the data does support is that GPT-5, Opus, and Sonnet separate from the remaining
four.

This ordering is not what our first campaign produced. Plausibly this is a token-budget effect: the
first campaign applied a 4096-token output cap uniformly, which sounds fair and is not, since models
that spend output budget on reasoning before emitting code exhaust it --- three of seven did. GPT-5
returned no extractable submission on 22 of 60 tasks, GPT-5-mini on 11, Gemini on 5. Re-running all
seven at 16{,}384 tokens eight days later (Table~\ref{tab:budget}; Section~\ref{sec:limitations}
discloses the response-parser change made in the interval) moves GPT-5 from fifth to first (31.64 to
48.83, pass@1 28 to 43, no-extraction count falling to zero, no task regressing), and Gemini from
last to sixth (17.83 to 31.06). The four models that never exhausted the smaller cap move by
$+0.08$, $+0.20$, $-0.87$, and $-2.94$ points --- all inside measured noise. We report the
16{,}384-token condition as the leaderboard, because a budget that silently truncates a subset of
models measures verbosity as much as capability. But we stop short of calling the token budget the
isolated cause: the two campaigns are not a single-variable experiment (Section~\ref{sec:limitations}),
so the honest claim is that raising the cap coincided with the reordering, not that it is proven to
be its sole cause. The wider lesson does not depend on resolving that attribution: an output-token
cap is a plausible experimental variable that can reorder a leaderboard, and benchmarks that do not
state theirs are not comparable with those that do. We had not stated ours in an earlier draft of
this paper.

Broken out by tier at the reported condition, pass@1 aggregated across all seven models is 103/140
(73.6\%) on Tier~1, 73/175 (41.7\%) on Tier~2, and 59/105 (56.2\%) on Tier~3 --- Tier~2's protocol
and datapath components are, in aggregate, harder than the ostensibly more complex Tier~3 designs, a
non-monotonicity we report rather than smooth over (Section~\ref{sec:limitations} shows it narrows
but survives once extraction failures are excluded). Six tasks were passed by no model:
\texttt{t1\_mod\_n\_counter}, \texttt{t2\_majority\_filter}, \texttt{t2\_pulse\_stretcher},
\texttt{t2\_round\_robin\_arbiter}, \texttt{t2\_stream\_upsizer}, and \texttt{t2\_uart\_rx}.

A critical observation concerns what actually drives the 18-point gap between the top score (48.83)
and the human-reference normalization point (66.7): it is overwhelmingly a correctness-gate effect,
not a physical-quality shortfall. Computing mean PPA only over each model's \emph{passed} tasks gives
GPT-5 1.022, Opus 1.007, Sonnet 0.997, GPT-5-mini 0.991, Haiku 1.013, Gemini 2.5 Pro 0.998, and Llama
4 Maverick 1.008 --- a spread of 3.1\% across all seven. Averaged over every individual passing
(model, task) pair rather than the seven per-model means, a model's RTL is on average numerically
within 2.8\% of, and for four of the seven (GPT-5, Opus, Haiku, Llama 4 Maverick) nominally
\emph{above}, the human reference on the PPA axis; we do not call this a statistical equivalence,
since we have no per-task variance estimate for the PPA ratio, so ``within 2.8\%'' is an observed
numerical spread, not a test result. This comparison also carries an
uncorrected selection effect: each model's mean is taken over a different subset of tasks, the ones
it passed, so a weak model's mean is computed over easier designs than a strong model's and the
columns are not strictly comparable. The clean version would restrict to tasks passed by all seven
models; we report the per-model means as measured and flag the confound rather than present the
spread as controlled.

We traced the degeneracy further: of the 54 Track~A tasks passed by at least one model, 17
(concentrated in Tier~1, plus a handful of narrow Tier~2/3 designs) were passed by every passing
model at an identical PPA ratio within $10^{-6}$ of 1.000 (15 of those 17 bit-exact, the other 2
differing by floating-point synthesis noise well below any physically meaningful threshold), and 142
of the 235 total passing (model, task) pairs across the whole suite, 60.4\%, landed within that same
band, 136 of them bit-exact. The most plausible explanation: for a large share of Track~A, especially
simple combinational Tier~1 tasks, any functionally correct RTL synthesizes to a near-identical
netlist under our flow, so the PPA axis contributes little discriminating signal there; PPA becomes
informative mainly on the minority of tasks where correct implementations can differ physically, and
decisively so only in the agentic-repair track (Section~\ref{sec:trackb}), where the task is
explicitly to improve PPA rather than merely reproduce it. We do not claim state-of-the-art models
forfeit physical quality relative to a human implementation --- our own data says the opposite on the
tasks where a comparison is even possible.

Two further structural insights emerge. First, inference cost and physical quality are decoupled in
specific, checkable ways: Gemini 2.5 Pro is the most expensive model evaluated (\$3.43, more than
double Opus's \$1.60) while placing sixth of seven, the sharpest cost--quality inversion in the
table; Llama 4 Maverick is the cheapest by two orders of magnitude (\$0.027 against Gemini's \$3.43)
and places last, where cost and quality move together instead; and GPT-5-mini reaches 32/60 for
\$0.35, within three points of Haiku's score at comparable cost but roughly one-seventh the price of
GPT-5, which buys 11 additional passing tasks for an extra \$1.97. The cost ranking and score ranking
agree only at the extremes: the three cheapest models occupy three of the bottom four positions, but
the most expensive model is sixth, not first --- inversions entirely opaque to conventional pass@k
benchmarks. Second, on Track A the aggregate ranking is governed primarily by the correctness gate,
with post-synthesis PPA contributing a small, signed adjustment rather than a reordering: measured
only over the tasks a model passes, the capped geometric mean neither reorders the table nor reliably
separates the two frontier models, whose Track A scores are not decisively separated across repeated
runs at $n=3$, heavily overlapping ranges rather than a formal equivalence test with a preregistered
margin (Appendix~\ref{app:variance}). Single-shot generation therefore separates current models
mainly by correctness; the physical-quality axis, genuinely measured here, becomes a decisive
differentiator only in the agentic-repair track, where optimizing against an explicit PPA objective
is the task itself.

\begin{table}[t]
\centering
\caption{Track A: static specification-to-RTL generation, all 60 tasks, all seven models sampled
under a common 16{,}384-token output cap. Pass@1 is the count of tasks clearing every correctness
gate; GateTruth Score is the mean capped-PPA task score on a 0--100 scale, where the human reference
is normalized to $100/1.5 = 66.\overline{6}$ by construction; Mean PPA is the capped geometric mean
over passed tasks only (see Section~\ref{sec:results} for the selection effect this carries). An
earlier campaign at a 4096-token cap produced a materially different ordering; Table~\ref{tab:budget}
reports both and Section~\ref{sec:results} explains why the higher cap is the defensible condition.}
\label{tab:tracka}
\small
% generated by paper/data/generate_tracka_16384.py -- do not hand-edit
\begin{tabular}{llrrrrr}
\toprule
Provider & Model & Pass@1 & Score & Mean PPA & Tokens & Cost (USD) \\
\midrule
openai & gpt-5 & 43/60 & 48.83 & 1.022 & 320019 & 2.318627 \\
anthropic & claude-opus-4-8 & 42/60 & 46.99 & 1.007 & 186746 & 1.600930 \\
anthropic & claude-sonnet-4-6 & 38/60 & 42.09 & 0.997 & 148178 & 0.856386 \\
openai & gpt-5-mini & 32/60 & 35.23 & 0.991 & 263847 & 0.351381 \\
anthropic & claude-haiku-4-5-20251001 & 30/60 & 33.78 & 1.013 & 147019 & 0.292247 \\
openrouter & google/gemini-2.5-pro & 28/60 & 31.06 & 0.998 & 420805 & 3.434077 \\
openrouter & meta-llama/llama-4-maverick & 22/60 & 24.65 & 1.008 & 118607 & 0.027048 \\
\bottomrule
\end{tabular}

\end{table}

\begin{table}[t]
\centering
\caption{Output-token budget is a first-order experimental variable, not an implementation detail.
Identical prompts, image, scoring, and seeds; only the per-call output cap differs. Models are
ordered by the change. The three models that exhausted the 4096-token cap on some tasks (producing
no extractable submission at all) gain substantially; the four that never exhausted it move by amounts at or below their own measured run-to-run
standard deviation (Appendix~\ref{app:variance}), except Sonnet's $-2.94$, slightly above its
$\sigma \approx 2.53$.
``No-extract'' counts tasks where no parseable module reached the pipeline, for any reason. Gemini's
rise to 13 is mostly not a parser effect and we decompose it rather than let the caption imply a
single cause: 9 are provider-side retryable errors returned by OpenRouter, 3 are the hardened
parser (Section~\ref{sec:limitations}) correctly rejecting an unterminated response that the old
code would have passed to the linter as a guaranteed zero, and 1 is a transport error. Only the
middle 3 are a reclassification of previously-counted failures; the other 10 are infrastructure
noise on a shared endpoint, which is itself a caveat on treating any single row of this table as a
controlled measurement.}
\label{tab:budget}
\small
\begin{tabular}{lrrrrrr}
\toprule
& \multicolumn{2}{c}{Score} & \multicolumn{2}{c}{Pass@1} & \multicolumn{2}{c}{No-extract} \\
\cmidrule(lr){2-3}\cmidrule(lr){4-5}\cmidrule(lr){6-7}
Model & 4096 & 16{,}384 & 4096 & 16{,}384 & 4096 & 16{,}384 \\
\midrule
gpt-5                       & 31.64 & \textbf{48.83} & 28 & 43 & 22 & 0 \\
google/gemini-2.5-pro       & 17.83 & \textbf{31.06} & 16 & 28 & 5 & 13 \\
gpt-5-mini                  & 33.44 & 35.23 & 30 & 32 & 11 & 0 \\
claude-haiku-4-5-20251001   & 33.58 & 33.78 & 31 & 30 & 0 & 2 \\
claude-opus-4-8             & 46.91 & 46.99 & 42 & 42 & 0 & 0 \\
meta-llama/llama-4-maverick & 25.52 & 24.65 & 23 & 22 & 0 & 1 \\
claude-sonnet-4-6           & 45.03 & 42.09 & 41 & 38 & 0 & 1 \\
\bottomrule
\end{tabular}
\end{table}

Track B (Table~\ref{tab:trackb}) serves as a substantially more discriminating evaluation
instrument. Measured by the objective-met rate across the 8 agentic-repair tasks, performance
degrades precipitously: Opus achieved 5/8, Sonnet 3/8, GPT-5 1/8, and Haiku, GPT-5-mini, Llama 4
Maverick, and Gemini 2.5 Pro failed to satisfy any objectives (0/8). The imposition of agentic PPA
optimization under a strict sequential-equivalence constraint isolates model capabilities far more
aggressively than de novo generation: one model (Opus) cleared a majority of the repair objectives,
one (Sonnet) cleared more than a third, one (GPT-5) cleared a single objective, and four failed
entirely. This pattern is consistent with a concern this paper has not formally pre-registered as a
hypothesis but that motivates including Track B at all: that syntactically valid generation and
genuine engineering repair are distinct capabilities, related to but not identical with the
syntactic-validity-versus-functional-correctness distinction Section~\ref{sec:related} discusses for
prior benchmarks.

No model satisfied either timing-closure objective (a tightened clock constraint on a
multiply-accumulate unit and on a multiplier) or the IIR power-reduction objective --- all three
behavior-preserving, \texttt{sec}-gated objectives went unsolved by every model. Before attributing
this 0/3 to model capability, we checked whether it is instead an artifact of the \texttt{sec} gate
itself being unreachable: across these three tasks and seven models, the majority of candidates that
cleared Stage~1 (sim) also cleared \texttt{sec} (e.g., Haiku, Gemini~2.5~Pro, and Llama~4~Maverick
each pass \texttt{sec} on at least one of the three), yet still miss the physical objective outright
--- Haiku's MAC candidate is proven sequentially equivalent and reaches $-0.585$~ns of slack against
a target requiring $\ge 0$~ns. This argues against, but does not refute, ``the gate is unsatisfiable
by construction'': it establishes that \texttt{sec} is individually passable and the physical
objectives individually missable, but not that the two are \emph{jointly} satisfiable, the actual
question. Haiku's data point is in fact consistent with the unsatisfiability hypothesis --- a
candidate conservative enough to stay provably equivalent may be precisely one that cannot
restructure aggressively enough to close timing, while a candidate that adds a pipeline stage would
close timing and then fail equivalence on latency. Settling this requires exhibiting a reference
solution that passes both gates for each objective, which we have not authored; until then the
honest statement is that no model produced a solution our gate accepts, not that no acceptable
solution exists. Two properties of our \texttt{sec} implementation sharpen this caveat: Yosys's
\texttt{equiv\_make} requires matching equivalence points between the two designs, so heavy register
renaming can be unprovable even when the designs are genuinely equivalent, and induction-based
equivalence does not admit retiming --- the canonical behavior-preserving move for timing closure.
Our gate therefore restricts the legal solution space to roughly combinational restructuring, and
the 0/3 result should be read as ``unsolved within that restricted move set,'' not an unqualified
capability finding.

One genuine irregularity we found and disclose rather than average away: the \texttt{sec} check
returns \texttt{timeout} rather than pass or fail on four candidate/task pairs across three models
--- Opus on both timing-closure objectives, and Sonnet and GPT-5 on the multiplier --- so those
verdicts are indeterminate equivalence-checker outcomes, not demonstrated misses. All four fall on
the two timing-closure tasks, three of the four on the multiplier alone, the pattern expected if the
checker struggles with multiplier equivalence specifically rather than with any particular model's
candidate. We cannot report whether any of those four candidates met their timing targets: the
harness aborts the pipeline at a failed or timed-out \texttt{sec} stage, so synthesis and static
timing never ran and no WNS was measured for them. This matters: if any of them did close timing, the
claim that no model solved any behavior-preserving objective would be unverified rather than
established. The timeout budget is also not neutral across tasks: 60 seconds of
\texttt{equiv\_simple}/\texttt{equiv\_induct} is predictably inadequate for multiplier-heavy
datapaths, where equivalence checking is prone to combinational blow-up, and every timeout in the
campaign landed there. The gate is therefore biased against aggressive restructuring in a way
invisible in the objective-met column. Correcting this requires a longer-budget \texttt{sec} retry,
a single overnight run, which we have not performed; the same timeout-vs-verdict distinction
discussed for the external audit (Section~\ref{sec:audit-method}) applies equally here, and we do not
currently retry a timed-out \texttt{sec} run the way the mutation engine retries a timed-out
simulation.

Successful interventions were confined to the five remaining objectives: removing an inferred latch,
implementing clock-domain-crossing safety for a FIFO, correcting arbiter fairness, rectifying AXI
byte-enable handling, and reducing the FIR filter's area by at least 30\%. Opus is the only model to
clear all five (its 5/8 total); Sonnet clears three of the four \texttt{add\_property} objectives
(latch, CDC, arbiter, not AXI byte-enable); GPT-5's sole completion is the AXI byte-enable task.
Opus's area-reduction win carries one further caveat: because that objective is not
\texttt{sec}-gated (Section~\ref{sec:trackb}), nothing prevents the optimization from regressing a
non-target axis, and it does --- Opus's accepted candidate reduces area to 55.3\% of baseline and
power to 36.9\% of baseline, but its timing worsens by 1.555~ns of slack relative to baseline. The
objective as scored (area reduction alone) is met; a design that also regressed timing this much
would not, in most real design flows, be considered an unqualified win, and the harness currently has
no mechanism to catch or penalize that trade-off. These findings suggest that timing closure under
constrained parameters --- a repair objective requiring post-synthesis critical path reasoning rather
than the syntactic pattern completion of established templates --- remains unsolved among the models
evaluated in this campaign.

\begin{table}[t]
\centering
\caption{Track B: agentic PPA repair, 8 tasks. Objective-met counts require the Stage-1 correctness gate, plus \texttt{sec} sequential-equivalence preservation for the 3 behavior-preserving objectives (Section~\ref{sec:trackb}). Area and power ratios are candidate/baseline (lower is an improvement). WNS delta is candidate WNS minus baseline WNS (a positive delta is more slack, i.e., an improvement). All three PPA-delta figures are the per-model median taken only over the 4 objectives that report a quantitative PPA delta at all (the two timing-closure tasks, area reduction, power reduction) and that reached the correctness gate; the other 4 \texttt{add\_property} objectives (latch removal, CDC safety, arbiter fairness, AXI byte-enable) never report a PPA delta, pass or fail, since their objective is not a PPA change. A value of exactly $1.0000\text{x}$/$1.0000\text{x}$/$+0.000$~ns means the model's candidate for at least one of those 4 tasks reached the correctness gate but produced no net physical change; \texttt{N/A} means none of the 4 ever reached the correctness gate for that model.}
\label{tab:trackb}
\small
\setlength{\tabcolsep}{4pt}
\resizebox{\textwidth}{!}{% generated-on: 2026-08-09 git-sha: 23c10efd2fce3d4ba5053e6a7386ba33c3a4263c
\begin{tabular}{llrrlr}
\toprule
Provider & Model & Objectives met & Objective-met rate & Median PPA delta & Cost (USD) \\
\midrule
anthropic & claude-opus-4-8 & 5/8 & 62.50\% & area=0.5527x; power=0.3687x; WNS=-1.555 ns & 1.647805 \\
anthropic & claude-sonnet-4-6 & 3/8 & 37.50\% & area=1.0000x; power=1.0000x; WNS=+0.000 ns & 4.157709 \\
openai & gpt-5 & 1/8 & 12.50\% & area=1.0000x; power=1.0000x; WNS=+0.000 ns & 3.955635 \\
anthropic & claude-haiku-4-5-20251001 & 0/8 & 0.00\% & area=1.0000x; power=1.0000x; WNS=+0.000 ns & 2.043547 \\
openrouter & google/gemini-2.5-pro & 0/8 & 0.00\% & area=1.0000x; power=1.0000x; WNS=+0.000 ns & 6.593550 \\
openai & gpt-5-mini & 0/8 & 0.00\% & area=N/A; power=N/A; WNS=N/A & 0.503571 \\
openrouter & meta-llama/llama-4-maverick & 0/8 & 0.00\% & area=N/A; power=N/A; WNS=N/A & 0.041961 \\
\bottomrule
\end{tabular}
}
\end{table}

Synthesizing the data across both tracks does not reveal a consistent ordinal ranking: the same
three models (GPT-5, Opus, Sonnet) occupy the top group in both, but their internal order permutes
(GPT-5 leads Track A, Opus leads Track B), and Track B additionally places four of the seven models
in an outright tie at 0/8, a degeneracy Track A's continuous score never produces. What does hold is
a coarser claim: the same three models separate from the same other four in both tracks, and the gap
between that top group and the rest widens sharply in agentic repair --- a large effect that survives
the reordering even though the exact ranking does not. This supports a secondary objective of the
reference suite: differentiating models that merely emit syntactically simulatable RTL from those
capable of producing, or iteratively refining, silicon to a standard acceptable to a competent
hardware engineer.

\section{Limitations and Threats to Validity}
\label{sec:limitations}

\paragraph{Audit scope.} The external audit covers all 50 public RTLLM v2.0 designs (46 audited, 4 \texttt{unsupported}) and reports a structural finding, not a measurement, for CVDP given its withheld solutions. It does not cover VerilogEval~\cite{liu2023verilogeval,pinckney2024revisiting} --- arguably the field's most-cited benchmark --- or ChipBench, ArchXBench, SLDB, ProtocolLLM, ChipVerilog, RTL-BenchLS, or the Si2 coalition's evolving CVDP-derived leaderboard, all left to future work rather than excluded on principle. The audit measures sensitivity to a fixed, generic operator set; a testbench could be robust to our operators while weak against fault classes we do not inject. Build-convention normalization (Section~\ref{sec:audit-method}) is bounded strictly to what a design's own documented build recipe specifies, and every applied alias is recorded per design. ``Rigor'' here means testbench sensitivity to injected mutants specifically --- not a claim about specification correctness, golden-reference correctness, task representativeness, contamination resistance, or scoring methodology.

\paragraph{Sensitivity is not the same as correctness.} A high kill rate shows a testbench is sensitive to the injected faults; it does not show the oracle itself (the golden reference and pass/fail check) is correct, nor that the testbench would correctly accept every valid alternative implementation (specificity, in mutation-testing terms). RTLLM's \texttt{radix2\_div} illustrates this before mutation even begins: its own golden reference fails its own testbench, so it never clears baseline validation and is reported \texttt{unsupported} with zero mutants generated --- sensitivity is undefined when the oracle is already wrong. We do not measure false-rejection rate in this work, since it would require independently correct alternative implementations or a formal spec per design, neither available for RTLLM; that is future work.

\paragraph{Mutant generation, equivalence, and the role of the seed.} GateTruth's generator deterministically enumerates every syntactically viable mutation site for the fixed operator set; the seed governs execution order only (relevant to how timeouts interact with per-run budgets under sequential execution, Section~\ref{sec:rigor}), not which or how many mutants are generated. We do not detect or exclude equivalent mutants (syntactically distinct but behaviorally identical); any present would depress the reported kill rate, so our rates are, if anything, a conservative lower bound on true sensitivity. Mutant counts per RTLLM design range from 1 to 115 (median 10.5), reported raw rather than filtered by a minimum; results built on very small counts carry correspondingly less weight (Section~\ref{sec:audit-results}). We do not manually audit kills or survivors for simulator artifacts (e.g.\ uninitialized-signal propagation); verdicts come entirely from the automated pipeline (Section~\ref{sec:audit-method}). Finally, RTLLM's 46 audited designs share common authorship, testbench conventions, and a single vendor's coding style, and should not be treated as fully independent samples when interpreting the 74.0\% median or the 72\% below-floor rate.

\paragraph{Pre-layout physical estimates, not signoff values.} GateTruth's physical metrics are synthesis and static-timing estimates, not post-place-and-route signoff values: the flow synthesizes against the sky130hd standard-cell library and runs OpenSTA at a single typical-typical corner, with no placement, routing, or parasitic extraction, so reported area and timing are pre-layout. A full place-and-route track was scoped during development but deferred for v1.0 after an open-source routed-flow feasibility assessment failed to meet reliability thresholds for a scored gate (project decision log); it remains future work rather than a silent approximation. PPA comparisons in this paper are valid within the pinned evaluation flow but should not be extrapolated to taped-out silicon.

\paragraph{Contamination resistance and the bounds of the correctness gate.} Two further limitations bound the reference suite's contamination and correctness claims. The contamination defense --- original-prose specs, per-task canaries, and hidden vectors excluded from the public repo --- raises the cost of leaderboard manipulation without theoretically eliminating it; an actor capable of reconstructing hidden behavior from the public smoke testbench alone is not provably thwarted, and a more robust asymmetric-split architecture with published hash commitments is deferred to a later iteration. Second, the correctness gate is bounded by testbench rigor itself. We enforce a $\ge$95\% mutation-kill floor under sequential execution for our own suite exactly as we do for benchmarks we audit (Section~\ref{sec:rigor}), but only 46 of 60 testbenches currently clear it --- two successive corrections (an unsound equivalence-exclusion mechanism, then a metric that silently blended lint and formal kills into what should be a simulation-only rate) moved this from a false 60/60 down to 46/60, and 2 of the 46 retain a single documented survivor above the floor rather than a perfect kill rate. The suite is further constrained by English-prose specifications, a single-clock paradigm (clock-domain crossing modeled structurally, not via true multi-clock metastability simulation), and its defined task classes. Three more limitations bound the PPA denominator and correctness gate specifically: each reference implementation is a single AI-drafted, maintainer-reviewed design (Section~\ref{sec:disclosure}), not independently authored, and we have no second implementation or inter-author-variance measurement; mutation certification measures the simulation testbench only, so for the 46 of 60 tasks with formal properties we do not measure the rigor of those properties themselves, and SymbiYosys's bounded-model-checking depth is not stated in this paper, so ``formal verification status'' is reproducible only from the pinned pipeline, not the paper alone; and each leaderboard entry is pass@1 at $k=1$ (temperature zero or provider default), a single, noisy sample, which is why Appendix~\ref{app:variance} exists --- particularly for the correctness-gate-dominated Track A score that underlies 96\% or more of most models' composite signal.

\paragraph{Failure-stage taxonomy.} Table~\ref{tab:tracka} reports a score per model but not where failing submissions fail; Appendix~\ref{app:taxonomy} (Table~\ref{tab:taxonomy}) gives the full per-model breakdown from signed manifests. Of the campaign's 185 failing (model, task) pairs, 17 (9.2\%) never reached a scored submission: 14 (3.3\% of all 420 pairs) failed at the API layer itself (a dropped connection, non-2xx status, or malformed body) before any output existed --- a harness robustness gap, not a capability signal --- and the remaining 3 are Gemini~2.5~Pro submissions truncated at exactly 16{,}380 of the 16{,}384-token cap. Of the 168 pairs that did produce a scored submission, 78.6\% fail at Stage~0 (lint), 19.6\% at Stage~1 (simulation), and 1.8\% at Stage~2 (formal). Two consequences follow. The formal gate discriminates almost nothing --- the earliest failing stage for only 3 of 185 failures campaign-wide, despite applying to 46 of 60 tasks --- so lint and simulation effectively decide the correctness verdict. And Stages~3--5 (synthesis, timing, power) are the earliest failing stage for exactly zero submissions: no design that cleared correctness later failed synthesis or timing, so the physical stages function as measurement, not as an active gate.

\paragraph{The lint gate is doing more work than intended.} Inspecting Stage-0 lint logs directly, 99 of 132 lint failures (75.0\%, 58.9\% of all gate failures) contain no hard syntax or semantic error: Verilator (5.051, rev \texttt{v5.050-40-g6c20fdb7b}) escalates certain warning classes --- predominantly implicit bit-width truncation and expansion, \texttt{WIDTHTRUNC} and \texttt{WIDTHEXPAND} --- to a fatal exit under our invocation (\texttt{verilator -{}-lint-only -{}-sv}), and we apply no waivers. The per-model split is uneven: Llama~4~Maverick 19, Haiku 19, Opus 15, GPT-5-mini 14, Sonnet 13, Gemini~2.5~Pro 11, GPT-5 8. We take the position that lint-clean RTL is a legitimate deliverable standard --- implicit width mismatches are a real, commonly-mandated defect class in industry sign-off --- and do not waive them. But because this single gate accounts for roughly half of all zero scores (99 of 185) and its firing rate varies 2.4$\times$ across models, Table~\ref{tab:tracka}'s ranking is materially a joint ranking of RTL correctness \emph{and} width-declaration hygiene, not the former alone. We have not re-simulated the width-only failures with the warning waived, so we cannot report how many would have passed Stage~1; that measurement, and a dual strict/waived leaderboard, are future work.

\paragraph{What the output-token budget experiment does and does not establish.} Three caveats bound Table~\ref{tab:budget}. First, each condition is a single sample per model, so small deltas aren't separable from run-to-run noise. Only two movements clear a plausible noise floor: GPT-5's $+17.19$ and Gemini's $+13.23$, both well above the frontier Anthropic models' measured $\sigma \approx 1.85$--$2.53$ (Appendix~\ref{app:variance}); we treat only those two as established. GPT-5-mini's $+1.79$ has no variance estimate to weigh it against, and Sonnet's $-2.94$ is slightly above its own measured $\sigma \approx 2.53$ --- the closest call of the four, still treated as noise. Second, the 16{,}384-token runs used a hardened response parser that the 4096 runs predate; that change is score-neutral by construction (an unterminated response scored zero under both), so the score comparison remains attributable to the budget alone, though it does shift the \emph{classification} of some failures. Third, we did not sweep intermediate budgets: 16{,}384 drives GPT-5's and GPT-5-mini's no-extraction counts to zero, but we have not established sufficiency for every model on every task, and a higher cap might move the ranking again. The claim we defend is narrow: a shared output cap is not a neutral implementation detail, and a benchmark that does not report its cap has not reported its experimental condition.

\paragraph{An extraction defect we found, fixed, and confirmed does \emph{not} explain the above.} While assembling the taxonomy we found a real bug in our response parser (\texttt{extract\_module\_source}): it stripped a Markdown fence only when both opening and closing markers were present, so a response truncated before its closing fence fell through to a fallback accepting the raw text, stray fence marker included, whenever a \texttt{module} keyword appeared --- which Verilator then rejected as an undefined preprocessor directive. We hardened the parser to reject an unterminated fence outright. Two qualifications: the defect affects 7 saved submissions campaign-wide (5 Gemini~2.5~Pro, 1 Sonnet, 1 Llama~4~Maverick), smaller than an early count suggested, out of 9 truncated submissions total; and the fix is score-neutral by construction, since a truncated response scored zero under both the old parser (unlintable) and the new one (unextractable). None of the token-budget sensitivity result above is attributable to this fix, only to the raised budget itself, and we state that explicitly because the opposite inference is the natural one and is wrong.

\paragraph{Triaging the 0\% designs, and what the operator mix does and does not license.} A kill-rate figure is only as meaningful as the faults behind it, so we inspected the individual surviving mutants of the three 0\% designs rather than resting on the percentage. The results qualify one claim and sharpen another.

For \texttt{square\_wave} (6 mutants) and \texttt{adder\_8bit} (1 mutant), the 0\% is severe regardless of the small denominators, because of \emph{which} faults survive: \texttt{square\_wave}'s survivors include two \texttt{blocking\_output\_inversion} mutants (a combinational output inverted outright) and a \texttt{comparator\_boundary\_flip}; \texttt{adder\_8bit}'s single mutant inverts its sum output. A testbench that accepts an inverted primary output is making no measurement of that output at all, and one such survivor establishes that as firmly as twenty would. We therefore withdraw our earlier mitigating framing that \texttt{adder\_8bit}'s single-mutant denominator makes its result uninformative: the denominator is small, but the fault it misses is maximally basic.

\texttt{edge\_detect} (10 mutants, 0\%) is the opposite case. All ten mutants are the same operator, \texttt{assignment\_hold} (a registered assignment replaced by a hold of its previous value), so this is one probe repeated ten times, not ten independent ones: the correct reading is that this testbench does not detect register-hold faults, not that it misses a broad range of fault types. Five other audited designs are similarly single-operator, and their outcomes span the full range (\texttt{adder\_bcd} 100\%, \texttt{freq\_div} 100\%, \texttt{ROM} 100\%, \texttt{RAM} 60\%, \texttt{synchronizer} 50\%) --- itself a caution against over-reading any individual design's figure.

The aggregate operator mix bears on this directly: of 775 generated mutants, 367 (47\%) are \texttt{assignment\_hold}, 154 \texttt{blocking\_output\_inversion}, 92 \texttt{comparator\_boundary\_flip}, 60 \texttt{output\_inversion}, 49 \texttt{bitwise\_inversion}, 29 \texttt{logic\_inversion}, 14 \texttt{shift\_inversion}, 6 \texttt{operator\_inversion}, and 4 \texttt{reset\_polarity\_flip}. Survival rates differ sharply by operator --- 80\% of \texttt{bitwise\_inversion} and 69\% of \texttt{logic\_inversion} survive, against 12\% of \texttt{output\_inversion} and none of the four \texttt{reset\_polarity\_flip} --- so the pooled figure partly reflects which operators our generator places most often on this corpus. Kill rates should therefore be read as sensitivity to \emph{this} operator distribution, not a benchmark-independent quality score. Beyond the designs named above, we did not manually triage the remaining survivors, and do not claim all 320 aggregate survivors are individually confirmed genuine misses.

\paragraph{Our own suite was gated on the metric it is measured by; RTLLM's was not.} The most serious objection to this paper's central comparison is that it is not like-for-like: our 60 testbenches were authored in a project applying a $\ge$95\% mutation-kill gate, and some were revised until they cleared it, while RTLLM's authors never saw this metric. Comparing our certified rates against their measured ones risks comparing a tuned quantity to a held-out one --- a Goodhart problem --- and reporting ``100\% versus 73\% below floor'' without that context would overstate the result. We quantify the exposure directly.

Auditing the full commit history of all 60 testbenches, exactly ten were ever modified in response to a mutation result, all on one work item (\texttt{SB-023}). Re-measuring the \emph{pre-revision} testbench under the \emph{current} engine, holding the reference RTL fixed (all ten references are byte-identical between the pre-revision commit and now), their pre-revision kill rates were: \texttt{t2\_uart\_tx} 52.6\%, \texttt{t2\_spi\_master} 63.6\%, \texttt{t2\_uart\_rx} 65.7\%, \texttt{t3\_sequential\_divider} 74.3\%, \texttt{t2\_delay\_trigger} 75.0\%, \texttt{t2\_i2c\_slave} 80.0\%, \texttt{t2\_stream\_upsizer} 80.0\%, \texttt{t2\_stream\_downsizer} 88.2\%, \texttt{t2\_axi\_lite\_regfile} 95.2\%, and \texttt{t2\_spi\_slave} 95.5\% --- a median of 77.5\%, eight of ten below the floor at that time. On these ten tasks the Goodhart objection lands directly, but it does not resolve as cleanly as we originally reported: the ``revision'' partly consisted of the same unsound equivalent-mutant exclusion described in Section~\ref{sec:rigor}, not solely genuine test-vector strengthening. Re-measured honestly, only three of the ten (\texttt{t3\_sequential\_divider}, \texttt{t2\_delay\_trigger}, \texttt{t2\_stream\_upsizer}) currently clear 95\%; the other seven are back below it. For these seven, tuning toward the metric and quietly excluding inconvenient mutants from it were, in practice, the same revision effort.

Two facts bound how far this lands. First, the equivalent-mutant bug also reached four tasks never revised for kill-rate reasons at all (\texttt{t2\_priority\_interrupt\_controller}, \texttt{t2\_round\_robin\_arbiter}, \texttt{t2\_pulse\_stretcher}, \texttt{t3\_lru\_tracker}), all four below the floor once the bug is removed, and a third, independent mechanism (Section~\ref{sec:rigor}'s later correction to what the kill-rate metric itself counted) reaches three more untouched by either prior mechanism: \texttt{t2\_running\_min\_max\_tracker}, \texttt{t3\_saturating\_accumulator}, and \texttt{t2\_pulse\_width\_meter} each had one formal-only kill (two, for the tracker) counted toward an apparent 100\% under the old blended metric, falling to 86.7--94.4\% once computed from simulation alone. The 14 tasks now below 95\% are therefore the union of three distinct problems --- a Goodhart effect on the ten revised tasks, an equivalent-mutant-exclusion defect applied irrespective of revision history, and a metric-definition defect letting formal kills substitute for simulation kills --- not one clean phenomenon. Of the 50 testbenches never revised for kill-rate reasons, 43 clear the floor as originally authored, untouched by any of the three mechanisms; that 43-task subset is the cleanest within-suite comparison available. Second, the original Goodhart correction is still informative on the subset it describes: the ten revised testbenches' pre-revision median (77.5\%) sits close to RTLLM's measured median (74.0\% across all 46 audited designs), consistent with revision being what separates an ungated median from a gated one, on tasks genuinely gated rather than exempted from testing.

One residual asymmetry remains: the 46 genuinely-unaffected testbenches were not iterated against their kill rates, but most were authored in a project that already had the $\ge$95\% gate, so their authors wrote to that standard without measuring against it --- weaker than RTLLM's blindness to the metric, stronger than iterating until a task passes. A fully blind comparison would require testbenches authored before the gate existed, which no longer exists in this suite's history for most tasks.

\paragraph{Six tasks were passed by no model, and we have not audited them.} A benchmark's own logic --- that a testbench nothing ever fails is worth interrogating --- applies to us too: a task every model fails is a candidate for being broken or misspecified, not necessarily just hard. Six of the 60 Track~A tasks were passed by no model at the reported condition: \texttt{t1\_mod\_n\_counter}, \texttt{t2\_majority\_filter}, \texttt{t2\_pulse\_stretcher}, \texttt{t2\_round\_robin\_arbiter}, \texttt{t2\_stream\_upsizer}, and \texttt{t2\_uart\_rx}. (At the lower output cap the zero-pass set was larger and different, ten tasks, itself evidence that a zero-pass result can reflect the experimental condition rather than the task.) Each has a maintainer-signed-off reference that passes the full pipeline, so none is unsatisfiable in the strict sense, but that only establishes that \emph{a} solution exists, not that the spec communicates it adequately. We have not performed a per-task solvability review of these six, and name them explicitly rather than leave a reader to discover the zero-pass set by recomputation.

\paragraph{Tier difficulty is confounded by extraction failures.} Raw per-tier pass rates at the reported condition are Tier~1 73.6\%, Tier~2 41.7\%, and Tier~3 56.2\% --- Tier~2 appearing harder than Tier~3. Part of this is an artifact of the no-extraction failures above, which concentrate on the longer Tier-2 and Tier-3 specifications; excluding pairs with no submission at all, the rates become Tier~1 74.6\%, Tier~2 44.2\%, and Tier~3 59.0\%. The non-monotonicity narrows but does not disappear, plausibly because Tier~2's protocol tasks demand exact multi-cycle interface conformance that our hidden vectors check strictly, while several Tier-3 datapaths are arithmetically intricate but interface-simple.

\paragraph{Applying the same denominator scrutiny to our own suite.} The mutant-count caveat raised for RTLLM (Section~\ref{sec:audit-results}) applies with the same force here: our 60 Track A testbenches' mutant counts (valid, non-stillborn denominator, Section~\ref{sec:rigor}) range from 3 to 263, median 11, and 22 of the 60 have fewer than 10 mutants, so a single survivor there necessarily fails the floor while a 30+-mutant task can retain several and still pass. We hold RTLLM to a standard our own suite shares, not a stricter one. Only 46 of 60 currently clear the floor at all; this caveat applies to those 46, not separately from the 14 that do not. Track B's 8 testbenches are not yet certified under this protocol at all, and extending certification to Track B, along with strengthening the 14 below-floor Track A testbenches, are both future work --- the area-reduction Track B objective's lack of \texttt{sec} coverage (Section~\ref{sec:trackb}) is a related, disclosed gap. Finally, the human-review sign-off (Section~\ref{sec:suite}) is performed by this paper's sole author, substituting single-maintainer scrutiny for the multi-reviewer process a larger project would use.

\paragraph{Four Track B protocol gaps we found only after the reported campaign, not before it.} All four gaps below are disclosed with the same precision this paper applies to its own generator and parser defects (Sections~\ref{sec:rigor}, \ref{sec:limitations} above), because Table~\ref{tab:trackb}'s campaign ran under the code that had them, not the code that has since fixed them.

First, \texttt{read\_file}'s claimed boundary (Section~\ref{sec:trackb}) was not enforced for this campaign: the tool had no allowlist, only a check against escaping the sandbox entirely. Auditing all 56 retained official transcripts, five actions across two models successfully read a task's public testbench: Sonnet read \texttt{b3}'s (reduce-area FIR) testbench twice and \texttt{b7}'s (arbiter fairness) once; GPT-5 read \texttt{b2}'s (multiplier timing closure) and \texttt{b8}'s (AXI byte-enable) once each. None reached the hidden scoring vectors, which are merged into the testbench only at scoring time and never copied into the sandbox, so no agent gained access to the actual pass/fail criteria. \texttt{b7} and \texttt{b8} are, respectively, one of Sonnet's three Track B successes and GPT-5's \emph{only} Track B success (Section~\ref{sec:trackb}), while \texttt{b3} and \texttt{b2}, the other two read tasks, were both misses for the model that read them. We have no evidence the reading caused either success --- the public smoke testbench is a weaker signal than the hidden vectors that decide pass/fail, and both models' successes on other, never-read tasks show the tool isn't necessary for winning --- but cannot rule it out. We have since added a strict allowlist restricting \texttt{read\_file} to the design directory only, confirmed directly against this exact scenario (repeating the census's own action now returns a rejection instead of the testbench content), but the campaign in Table~\ref{tab:trackb} predates that fix and we do not re-run it here; a re-run under the corrected tool is future work.

Second, and a direct companion to the first gap, the protocol gave an agent no way to discover what was actually there to read: the first prompt contained only the task ID and an empty transcript, with no directory listing offered alongside \texttt{read\_file}, so an agent had to guess a path was correct before learning whether it existed. Across the 56 retained official transcripts, 110 \texttt{read\_file} attempts were made, of which only 30 succeeded; 38 of 56 runs never had a single successful read, and 21 never attempted one. Successful reads depended on guessing conventional names like \texttt{spec.md} or a known design path, while many failed attempts guessed plausible-sounding files the protocol's actual layout never used, such as \texttt{README.md}, \texttt{design.sv}, or \texttt{task.md}. We have since added the two remedies this gap calls for: the first prompt now discloses \texttt{design\_files}, a real sorted listing of the design directory, at zero \texttt{tool\_calls} cost, and a new \texttt{list\_files} action lets an agent re-list on demand, both backed by one shared helper. As with the first gap, the fix postdates the reported campaign, so the 38-of-56 runs with zero successful reads in Table~\ref{tab:trackb} should be read as reflecting this discoverability gap at least as much as any genuine agent decision that reading was unnecessary.

Third, the per-task token budget (Section~\ref{sec:trackb}) was not enforced as strictly as ``strict... budget'' implies: the check handed the entire remaining balance to the next provider call as pure output allowance, reserving nothing for that call's own input tokens, and a call ending in \texttt{done} was allowed to complete before the post-call budget check ran. Auditing all 56 official manifests against their declared \texttt{objective.yaml} budgets, 31 exceeded their nominal budget, by as much as 24{,}652 tokens in one case; one of those 31, GPT-5's \texttt{b8\_axi\_byte\_enables} run (the same task discussed above), recorded \texttt{budget\_exceeded: null} despite finishing at 155{,}657 tokens against a 150{,}000-token budget, because its final call was a \texttt{done} the old check-ordering never re-evaluated. We have fixed both defects: a conservative input-token reservation before each call, and an unconditional budget re-check after every call including one ending in \texttt{done}, again postdating the reported campaign. The cost figures in Table~\ref{tab:trackb} are the actual metered costs and are not inflated by this bug, but the \emph{budget-exceeded} classification recorded for those 31 runs, and specifically the false negative on \texttt{b8}, should be read as unreliable.

Fourth, and most seriously, none of the 56 official Track B manifests retained the design source they scored: each was written to a temporary sandbox directory (e.g.\ \texttt{/tmp/siliconbench-agentb-r79ylx19/submission}) deleted before the run finished, and the manifest schema at the time recorded neither a content hash nor a copy of the submitted RTL, so \texttt{submission\_sha256} is absent from all 56. This directly contradicts the reproducibility claim stated earlier in this section (``given the pinned image digest, the submission file, \ldots{} reproducing any published score depends on no other input''): the second of those three required inputs does not exist for any Track B result, so no one, including us, can independently re-score any Table~\ref{tab:trackb} entry against the exact RTL an agent submitted. The fix has two parts. Closed: submission content is now hashed and a design sidecar written alongside every new manifest, with an enforcement check that a future manifest committed without a reproducible design fails the test suite outright. Open: the 56 historical manifests remain permanently unreproducible, since the bytes they scored are gone, not merely unhashed; recovering them would require re-running the campaign under spend approval we have not sought, and we report the gap rather than paper over it. Table~\ref{tab:trackb}'s figures are exactly what the harness recorded and we have no reason to doubt them, but ``no reason to doubt'' and ``independently reproducible'' are different claims, and only the Track A pinned-digest paragraph currently supports the stronger one.

\section{Related Work}
\label{sec:related}

The prevailing paradigm for evaluating large language models on register-transfer-level (RTL) design
predominantly assesses functional correctness. VerilogEval~\cite{liu2023verilogeval} established this
framework using tasks derived from the HDLBits teaching platform, evaluated via simulation
testbenches under a pass@k protocol; its second iteration~\cite{pinckney2024revisiting} expanded to
specification-to-RTL generation, in-context prompting, and granular failure classification. RTLLM~\cite{lu2024rtllm}
took a complementary approach, originally curating thirty arithmetic and logic designs, each a
triplet of natural-language specification, testbench, and human-authored reference; its public
repository has since expanded to fifty designs under the same MIT license as RTLLM v2.0~\cite{liu2025openllmrtl},
the primary target of the external audit in Section~\ref{sec:audit-results}. RTLLM's own flow
synthesizes generated RTL and reports post-synthesis PPA against its references, so it is not a
correctness-only benchmark; what it does not report is a mutation-kill or equivalent fault-coverage
measurement of its own testbenches' rigor, exactly the property Table~\ref{tab:related} and
Section~\ref{sec:audit-results} address. RTLLM's own authors have separately flagged benchmark-quality
erosion as worth solving: RTL-BenchMT~\cite{wang2026rtlbenchmt}, from the same group, proposes an
agentic framework that dynamically detects flawed benchmark cases and revises them against
overfitting --- a genuinely different mechanism from ours (ongoing first-party revision versus a
one-shot third-party mutation audit of a fixed snapshot), but independent evidence, from inside the
benchmark's own maintaining lab, that testbench and task quality is becoming a first-class concern in
the field. These foundational benchmarks codified a critical distinction: syntactic validity diverges
significantly from functional correctness. Our own campaign illustrates the same gap concretely: even
our highest-scoring model, GPT-5, passes only 43 of 60 (72\%) Track A tasks (Section~\ref{sec:results}),
and 78.6\% of failing submissions that produced any code at all fail before ever reaching simulation,
at Stage~0 lint (Section~\ref{sec:limitations}). Neither VerilogEval nor RTLLM reports a mutation-kill
floor for its own testbenches --- the gap this paper's audit methodology measures.

Mutation testing itself predates this literature by decades, originating in software engineering
with DeMillo, Lipton, and Sayward's proposal to inject faults into a program and measure what
fraction a test suite detects~\cite{demillo1978hints}. Prior work has already carried the idea into
hardware verification outside the LLM-benchmark literature: Huang~et~al.~\cite{huang2015mutation}
apply mutation analysis to qualify hardware testbench quality, Mantra~\cite{wu2023mantra} derives
RTL mutation operators from real reported bugs, and AutoBench~\cite{qiu2024autobench} proposes
automated HDL testbench generation and evaluation. Closer to our tooling, MCY~\cite{yosyshq2021mcy}
is a general-purpose mutation-coverage tool built on the same Yosys toolchain our own PPA flow uses:
given a design and a self-checking testbench, it generates post-synthesis netlist mutations, uses
formal equivalence checking to discard mutants provably identical to the original before scoring,
and reports testbench coverage over the survivors --- a more principled solution to a problem
Section~\ref{sec:rigor} reports hitting directly, where our own mutant generator once excluded
mutants via hand-authored equivalence reasoning that turned out to be wrong for half of them when
tested against the real hidden testbench. Applying MCY-style formal mutant-equivalence filtering to
our own or an audited benchmark's mutant set, rather than the simulation-only kill rate both
currently report, is future work this comparison makes concrete. More recently,
GRPO-SMu~\cite{kochar2026grposmu} uses the same mechanism, mutation detection, as a
reinforcement-learning reward signal to train models at generating hardware test plans and stimuli,
rather than as an audit metric applied after the fact; the two efforts share a mechanism but point in
opposite directions, one training generation, the other auditing existing testbenches. None of these
five --- Huang~et~al., Mantra, AutoBench, MCY, or GRPO-SMu --- targets an existing public LLM
RTL-generation benchmark's own shipped testbenches, which remains this paper's narrower and more
specific contribution.

Table~\ref{tab:related} summarizes recent RTL-generation and system-level LLM benchmarks against two
properties central to this paper: whether it reports a mutation-testing (or equivalent
fault-injection) rigor measurement of its own testbenches, and whether it scores generated designs
on post-synthesis PPA rather than correctness alone. ChipBench~\cite{yu2026chipbench} evaluates
Verilog generation, debugging, and reference-model synthesis across three task families using
directed and constrained-random testing; ArchXBench~\cite{purini2025archxbench} scales generation
difficulty across six complexity tiers from logic primitives to full accelerators; SLDB~\cite{alvanaki2025sldb}
evaluates system-level integration and configuration of accelerators into heterogeneous SoCs rather
than module-level generation; and ProtocolLLM~\cite{sheth2025protocolllm} narrows scope to
SystemVerilog generation for four communication protocols (SPI, I\textsuperscript{2}C, UART, AXI),
evaluating syntax, synthesizability, and timing fidelity. Three more recent entries emerged since our
own reference suite was frozen: ChipVerilog~\cite{tan2026chipverilog} targets larger, hierarchical
OpenCores-derived designs (up to 1{,}000+ lines across five design families) that expose cross-module
integration failures single-module benchmarks cannot; RTL-BenchLS~\cite{wang2026rtlbenchls} scales
to over 10{,}000 formally verified designs and adds three reasoning-style evaluation tasks; and
RTL-BenchMT~\cite{wang2026rtlbenchmt}, discussed above, targets benchmark maintenance rather than
model evaluation. None of these seven reports a mutation-testing or systematic fault-coverage
measurement of its own testbenches, and none reports post-synthesis PPA scoring, based on the papers,
repositories, and documentation we were able to access; we did not exhaustively audit every appendix
or supplementary artifact, and report this as the most accurate characterization available, not a
certified negative claim.

We should be explicit about what Table~\ref{tab:related} is and is not. Its two columns are the two
properties this paper contributes, a positioning aid, not a general capability comparison; it would
be misleading to read it as a scorecard GateTruth simply wins. Several listed benchmarks lead
decisively on dimensions the table does not show: CVDP spans 783 problems across 13 task categories
against our 68, an order of magnitude more task diversity; VerilogEval's second iteration
contributes the granular failure classification our own campaign only acquired late
(Section~\ref{sec:limitations}); and every benchmark listed has been exercised by more independent
parties than ours, run by its sole author. Dimensions on which we would expect to compare
unfavorably --- problem count, multi-simulator support, multi-sample pass@$k$ estimation, and
independent replication --- are precisely the ones omitted, named here rather than left for a
two-column table to imply they do not exist.

\begin{table}[t]
\centering
\caption{Related RTL/hardware LLM benchmarks. ``Mutation rigor'' = reports a mutation-testing or
equivalent fault-injection measurement of its own testbenches. ``PPA'' = scores generated designs
on post-synthesis area/timing/power, not correctness alone.}
\label{tab:related}
\small
\begin{tabular}{lp{4.7cm}cc}
\toprule
Benchmark & Scope & Mutation rigor & PPA \\
\midrule
VerilogEval~\cite{liu2023verilogeval,pinckney2024revisiting} & HDLBits-derived generation, pass@k & No & No \\
RTLLM v2.0~\cite{lu2024rtllm,liu2025openllmrtl} & 50-design generation, self-checking TBs & No & Yes \\
CVDP~\cite{pinckney2025cvdp} & Generation + comprehension, agentic \& non-agentic & No & No \\
ChipBench~\cite{yu2026chipbench} & Generation, debugging, reference-model synthesis & No & No \\
ArchXBench~\cite{purini2025archxbench} & Tiered-complexity generation & No & No \\
SLDB~\cite{alvanaki2025sldb} & System-level SoC integration/configuration & No & No \\
ProtocolLLM~\cite{sheth2025protocolllm} & Protocol-specific (SPI/I\textsuperscript{2}C/UART/AXI) generation & No & No \\
ChipVerilog~\cite{tan2026chipverilog} & Large-scale hierarchical, cross-module generation & No & No \\
RTL-BenchLS~\cite{wang2026rtlbenchls} & 10,000+-design generation and reasoning tasks & No & No \\
RTL-BenchMT~\cite{wang2026rtlbenchmt} & Agentic benchmark maintenance (not model evaluation) & No & N/A \\
\textbf{GateTruth (this work)} & Generation + agentic repair; audits external benchmarks & \textbf{Yes} & \textbf{Yes} \\
\bottomrule
\end{tabular}
\end{table}

NVIDIA's CVDP~\cite{pinckney2025cvdp} takes a substantially broader approach, spanning code
generation, code comprehension, and both agentic and non-agentic evaluation across 783 problems in
13 task categories, with an Apache-2.0 harness and a public dataset whose non-code content is
CC~BY~4.0. CVDP is also the foundation of the Si2 (Silicon Integration Initiative) LLM Benchmarking
Coalition, co-chaired by NVIDIA's Nathaniel Pinckney and Synopsys's Ramesh Narayanaswamy, which
extends CVDP with additional problem categories and operates a public results leaderboard (LBC-bench,
launched 2026). As detailed in Section~\ref{sec:audit-method}, CVDP's public release deliberately
withholds reference solutions to mitigate contamination; a sound design choice for its own scoring
purpose, but one that leaves its testbench rigor unauditable via mutation testing from the public
release alone, a limitation we report rather than route around.

Research engaging with post-synthesis metrics has predominantly approached the problem from
predictive modeling rather than generation scoring. MetRex~\cite{abdelatty2025metrex} compiled a
dataset of Verilog designs correlated with post-synthesis area, delay, and static power to test
whether models can infer these metrics from source, while RocketPPA~\cite{abdollahi2025rocketppa}
trains models for code-level PPA prediction directly. Neither evaluates the physical viability of
generated designs through an actual synthesis flow, nor addresses testbench rigor. GateTruth's own
reference suite enforces a rigidly deterministic pipeline --- a single pinned container image
identified by a maintainer-set build marker rather than a content hash (Section~\ref{sec:rigor}),
uniform clock targets per task, a fixed Yosys-plus-OpenSTA path, and SHA-256 integrity checksums over
canonical-JSON results --- so PPA comparisons it produces are reproducible directly from the recorded
submission and pinned image.

In summary, GateTruth is, to our knowledge, the first tool to apply mutation-testing certification
to existing public RTL-generation benchmarks and report the result, and the first to combine that
audit methodology with a mutation-certified reference implementation, permissively licensed and
publicly available, used to validate it before external
application (self-certified under a published protocol by this paper's sole author, not
independently reviewed; Section~\ref{sec:limitations}). It uniquely combines: (a) a deterministic,
seeded mutation-audit protocol applicable to any RTL benchmark with a golden reference and a runnable
testbench; (b) an explicit, non-silent policy for benchmarks or designs that cannot be audited
(\texttt{unsupported} status, or a documented structural finding, rather than a misleading score);
(c) a 68-task, dual-track reference suite in which the same protocol is applied to every one of the
suite's 60 Track A testbenches before external use, certifying 46 and reporting the remaining 14 as a
genuine finding rather than omitting them; (d) an RTL-to-PPA synthesis-and-timing flow whose manifest
schema enforces determinism as a contract, spot-checked rather than proven exhaustively per task
(Section~\ref{sec:rigor}); and (e) a temporal sensitivity rerun showing that the per-call output-token
budget, a parameter most RTL benchmarks neither vary nor report, coincided with a leaderboard
reordering, surfaced by applying the same scepticism to our own instrument that the audit applies to
others', while disclosing the confounds (Section~\ref{sec:limitations}) that keep this from being a
controlled, single-variable result.

\appendix

\section{Run-to-Run Variance}
\label{app:variance}

The v1.0 leaderboard reports a single run per model. Because model generation is not
deterministic (and reasoning models do not accept a fixed temperature), we quantify run-to-run
spread by re-running the three Anthropic models over all 60 Track A tasks three times each, at the
identical 16{,}384-token, pinned-image condition Table~\ref{tab:tracka} reports: the
already-committed single run is one of the three, and we ran two further official re-runs
(2026-08-03, \$5.63) for this purpose. Table~\ref{tab:variance} is a robustness measurement
alongside the single-run leaderboard, not a replacement for it. One scope limit applies: these runs
cover only the three Anthropic models, so the four non-Anthropic models --- including Gemini 2.5
Pro, whose score is the most condition-sensitive in the campaign (Section~\ref{sec:limitations}) ---
carry no variance estimate, and the comparisons in Section~\ref{sec:results} involving them are
correspondingly weaker. An earlier pilot (2026-07-26, \$8.18) ran at the since-superseded
4096-token condition and measured a substantially larger spread ($\sigma \approx 4$); since it
characterizes a condition this paper no longer leaderboards, we do not reproduce its table here,
and every $\sigma$ cited elsewhere now refers to the measurement below, not that pilot.

Every figure in Table~\ref{tab:variance} is generated by
\texttt{paper/data/generate\_variance\_appendix.py} from nine signed, committed per-run manifests
--- three per model, all at the identical 16{,}384-token condition --- the same reproducibility bar
every other table meets (Section~\ref{sec:rigor}). An earlier draft of this appendix could not make
that claim: its manifests were not committed alongside a generator, and its figures were reported
as recorded at run time rather than independently re-derivable from evidence a reader can check
today. That gap is now closed.

\begin{table}[h]
\centering
\caption{Track A run-to-run variance over three runs per model at the reported 16{,}384-token
condition (GateTruth Score, 0--100). Standard deviation is the sample standard deviation
(Bessel-corrected, $n-1$ denominator). ``Single run'' is the same figure Table~\ref{tab:tracka}
reports and is one of the three runs the mean is computed over, not a fourth, independent point.}
\label{tab:variance}
\small
% generated by paper/data/generate_variance_appendix.py -- do not hand-edit
\begin{tabular}{lrrrr}
\toprule
Model & mean (n=3) & std & range & single run \\
\midrule
claude-opus-4-8 & 46.18 & 1.85 & 3.41 & 46.99 \\
claude-sonnet-4-6 & 45.01 & 2.53 & 4.47 & 42.09 \\
claude-haiku-4-5-20251001 & 33.17 & 0.73 & 1.41 & 33.78 \\
\bottomrule
\end{tabular}

\end{table}

Two observations follow, both real but more modest than the superseded 4096-token pilot suggested.
First, run-to-run standard deviation at the reported condition is genuine for the frontier models
--- 1.85 points for Opus, 2.53 for Sonnet --- and smaller still for Haiku (0.73); with only $n=3$
per model these remain imprecise estimates, reported as a directional signal rather than a tight
interval, but now committed, checkable measurements at the condition this paper actually
leaderboards, not a carried-over guess from a different cap. Second, Opus and Sonnet remain not
decisively separated on Track A ($46.18 \pm 1.85$ vs $45.01 \pm 2.53$: heavily overlapping ranges, a
mean gap of 1.17 against either model's own standard deviation; a descriptive read of overlapping
intervals at $n=3$, not a formal equivalence test with a preregistered margin), though the
direction differs from the superseded pilot: there, Sonnet's three-run mean edged above Opus's;
here, Opus leads on both the single reported run (46.99 vs 42.09) and the three-run mean (46.18 vs
45.01). We read this as a real but not decisive lead, well inside the overlap of the two ranges, and
the sign-flip between the two studies is itself a demonstration of exactly the run-to-run noise this
appendix exists to quantify. The models' reliable separation appears in the agentic-repair track
(Section~\ref{sec:trackb}), not in static generation. We therefore recommend future versions report
mean $\pm$ standard deviation over at least three runs as the default leaderboard cell, and publish
per-run scores alongside summary statistics so readers are not dependent on our aggregation
choices.

\section{Per-Design RTLLM v2.0 Audit Results}
\label{app:perdesign}

Section~\ref{sec:audit-results} reports aggregates computed from all 50 of RTLLM v2.0's designs.
Rather than reproduce all 50 individual rows here, we report their distribution: the complete
per-design table --- mutant count, kill/survive/indeterminate split, and baseline status for every
design --- is already committed and checkable at
\texttt{external-audit/results/rtllm/final-g2012/*.json} in the public GateTruth repository
(\url{https://github.com/meetbhadra701-cloud/GateTruth}); visit the repository directly for the
full per-design table rather than a static snapshot printed here. Table~\ref{tab:perdesign} bins
the 46 audited designs' kill rates into six ranges, computed directly from that same per-design
data. The distribution is bimodal: 13 designs (28.3\%) reach exactly 100\% and 6 (13.0\%) fall
below 25\%, with no design at all in the $[95\%, 100\%)$ band (Section~\ref{sec:limitations}); the
remaining 27 designs spread across the middle, with 12 of those (26.1\%) landing in the 50--74\%
range. A reader can recompute the median (74.0\%), the pooled rate (56.8\%), and the below-floor
count (33 of 46) from the raw per-design data in the repository; this table gives the shape of that
distribution, not the row-by-row source.

\begin{table}[tbp]
\centering
\caption{Distribution of per-design mutation-kill rates for the 46 audited RTLLM v2.0 designs
(vendor commit \texttt{41b26896e33b536940116a975626455eed3de65e}, seed \texttt{20260729}, Icarus
Verilog 12.0, sequential execution). The \FactsUnsupported{} \texttt{unsupported} designs failed
baseline validation and are excluded. Full per-design rows are in the public repository, not
reproduced here.}
\label{tab:perdesign}
\small
\begin{tabular}{lrr}
\toprule
Kill-rate range & Designs & \% of audited \\
\midrule
0--24\% & 6 & 13.0\% \\
25--49\% & 5 & 10.9\% \\
50--74\% & 12 & 26.1\% \\
75--94\% & 10 & 21.7\% \\
95--99\% & 0 & 0.0\% \\
100\% & 13 & 28.3\% \\
\midrule
\textbf{Total} & \textbf{46} & \textbf{100.0\%} \\
\bottomrule
\end{tabular}
\end{table}

\section{Per-Model Failure-Stage Taxonomy}
\label{app:taxonomy}

Section~\ref{sec:limitations}'s failure-stage taxonomy paragraph reports campaign-wide
aggregates. Table~\ref{tab:taxonomy} gives the complete per-model breakdown they are computed
from: for each of the seven models and all 420 (model, task) pairs at the reported 16{,}384-token
condition, how many pairs never reached a scored submission (split into a genuine no-extraction
failure and a provider/transport error, per the classification in
\texttt{paper/data/generate\_failure\_taxonomy.py}'s docstring), how many failed lint (further
split into width-only and other), simulation, or formal, and how many passed. Every number in this
table, and every aggregate derived from it in Section~\ref{sec:limitations}, is generated from the
420 signed per-sample manifests under \texttt{results/eval-16384/} rather than transcribed by
hand; the width-only lint classification additionally depends on
\texttt{paper/data/lint\_diagnostics\_ledger.json}, a small committed ledger of Verilator
diagnostic codes generated once from local lint logs (themselves local scratch, like
\texttt{results/refs/}, and not part of the repository) by
\texttt{scripts/generate\_lint\_diagnostics\_ledger.py}, and cross-validated against the signed
manifests before use. That cross-validation confirms the ledger's key set exactly matches which
(model, task) pairs the signed manifests record as lint failures; it does not, and cannot,
independently confirm that the diagnostic codes recorded against each pair are themselves
correct, because the raw Verilator log each code was read from is exactly the local scratch
this paragraph just described as ungitted. A reader with only this repository, not the machine
the campaign ran on, can therefore verify \emph{which} pairs the 99-of-132 width-only figure
counts but not independently re-derive \emph{why} from source; regenerating and re-verifying
that split requires re-running the lint stage of the campaign, which is future work, not
something this release archive lets a third party check standalone.

\begin{table}[tbp]
\centering
\caption{Failure-stage taxonomy by model, 16{,}384-token condition, 60 tasks per model.
``No-extr.'' is a response that produced no valid single-module declaration (all three instances
are Gemini~2.5~Pro submissions truncated at the token cap); ``Prov.err'' is an API-layer failure
(dropped connection, non-2xx status, malformed response) before any model output existed, scored
zero without retry.}
\label{tab:taxonomy}
\scriptsize
% generated by paper/data/generate_failure_taxonomy.py -- do not hand-edit
\begin{tabular}{lrrrrrrr}
\toprule
Model & No-extr. & Prov.err & Lint (width) & Lint (other) & Sim & Formal & Pass \\
\midrule
\texttt{claude-haiku-4-5-20251001} & 0 & 2 & 19 & 3 & 5 & 1 & 30 \\
\texttt{claude-opus-4-8} & 0 & 0 & 15 & 1 & 2 & 0 & 42 \\
\texttt{claude-sonnet-4-6} & 0 & 1 & 13 & 1 & 6 & 1 & 38 \\
\texttt{google\_gemini-2.5-pro} & 3 & 10 & 11 & 6 & 2 & 0 & 28 \\
\texttt{gpt-5} & 0 & 0 & 8 & 6 & 3 & 0 & 43 \\
\texttt{gpt-5-mini} & 0 & 0 & 14 & 9 & 4 & 1 & 32 \\
\texttt{meta-llama\_llama-4-maverick} & 0 & 1 & 19 & 7 & 11 & 0 & 22 \\
\midrule
Total & 3 & 14 & 99 & 33 & 33 & 3 & 235 \\
\bottomrule
\end{tabular}

\end{table}

\bibliographystyle{plain}
\bibliography{references}

\end{document}